\documentclass{desyproc}
\usepackage{slashed}
\newcommand{\eps}{\epsilon}
\newcommand{\D}[1]{\frac{d^{D} #1}{(2\pi)^{D}}}
\newcommand{\im}{{\rm Im}}
\newcommand{\IP}{\mbox{I}\!\mbox{P}}
\begin{document}
\title{ \vspace*{-2cm}
        \begin{flushright}
        {\footnotesize MZ-TH/09-21}
\end{flushright}
\vspace{0.5cm}Radiative corrections to top quark decays}
\author{{\slshape A. Kadeer$^{1,2}$, J.G. K\"orner$^1$}\\[1ex]
$^1$Institut f\"ur Physik, Johannes-Gutenberg-Universit\"at,
Staudinger Weg 7\\
D-55099 Mainz, Germany\\[1ex]
$^2$ Deutsches Elektronen-Synchrotron\\
D-22607 Hamburg, Germany
 }

\acronym{HQP08} 
\maketitle

\begin{abstract}
We provide a pedagogical introduction to the subject of Standard Model decays
of unpolarized top quarks into unpolarized and polarized $W$-bosons
including their QCD and electroweak radiative corrections. 
\end{abstract}

\section{Introductory remarks}
These lectures held by one of us (JGK) at the II Helmholtz International 
Summer 
School on Heavy Quark Physics in Dubna, Russia (August 11 - 21 2008) are 
meant as pedagogical lectures aimed at the level of the
audience which, on the participants' side, was composed of graduate students 
with a few postdoctoral students mixed in. We give many details on the Born
level calculation of rates and angular decay distributions which can be
profitably used in the higher order radiative correction calculations.
The material collected in the write-up of the lectures given by one of us at
the International School on Heavy Quark Physics in Dubna, Russia 
(27 May - 5 Jun 2002) \cite{Korner:2003zq} covering similar topics will not 
always be repeated. In addition to the review \cite{Korner:2003zq} we very
much recommend the excellent reviews on top quark physics in 
\cite{Kuhn:1996ug,Chakraborty:2003iw,Bernreuther:2008ju,Wagner:2008zz,
Incandela:2009pf}. One of the main aim of 
these lectures is to illustrate advanced loop techniques in simple Born term
 settings. We begin by listing the basic properties of the Standard Model (SM) 
top quark and its SM decay features.
\subsection{ {\bf Mass of the top quark}} 
In our numerical calculations we always take the top quark mass to be 
$m_{t}=175$\,GeV.
The latest Tevatron combination
is $ m_{t}=173.1\pm 0.6(stat.)\pm 1.1(syst.)$\,GeV \cite{Group:2008nq}. Since
all our results are in closed analytical form any other value of the top 
quark mass can be used as input in these formulas.

There have been suggestions for indirect measurements of the top quark mass
through the measurements of dynamic quantities that depend on the value of
the top quark mass. For example, the SM $(t\bar{t})$-production rate at e.g. 
hadron colliders
is sensitive to the value of the top quark mass (in particular at the 
Tevatron II) and thus the $(t\bar{t})$-production rate could  be used to
``measure'' the top quark mass. Another possibility is to
accurately measure the longitudinal and transverse-minus helicity decay rates
of the top quark. The ratio of the two helicity rates is well suited for an 
indirect determination of the top quark
mass since the ratio depends quadratically on the top quark 
mass, i.e. $\Gamma_{L}/\Gamma_{-}\sim m_{t}^{2}/m_{W}^{2}$. One should,
however, always take into account radiative corrections in such indirect
top quark mass measurements. For example, in the latter case 
the NLO QCD and electroweak radiative corrections have different effects on 
the two partial helicity rates which lead to a $3.6\%$ upward shift in the 
helicity rate ratio $\Gamma_{L}/\Gamma_{-}$ for top quark masses around
175 GeV 
\cite{Fischer:2001gp,Do:2002ky}.

In a third method one measures the mean distance that $b$-hadrons from
$(t \bar{t})$-events travel before they decay \cite{Hill:2005zy}. The mean 
distance is obviously correlated with the value of the top quark mass.
Needless to say that all these indirect top quark
mass measurements crucially depend on the assumed correctness of the SM.

\subsection {Top quark decays before it can hadronize} 

Singly produced top quarks in hadronic collisions are produced by weak
interactions and are almost 100\%
polarized.
The top quark retains its polarization which it has at birth when it decays. 
The standard
argument is that the life time of the top quark
($\tau_{t}\cong 4.6 \times 10^{-25}s$) is shorter than the hadronization
time which is characterized by the inverse of the nonperturbative scale of QCD,
i.e.$\Lambda^{-1}_{QCD}\cong 10^{-23}s$. 

However, one can do better as pointed
out in \cite{Grossman:2008qh} who extended earlier work on depolarization
effects in the bottom sector~\cite{close92,Falk:1993rf}. Consider a polarized top quark which picks up 
a s-wave light antiquark of opposite spin direction. This state will be a 
coherent superposition of the spin 0 and spin 1 mesonic ground states as 
follows
\begin{equation}
t(\uparrow)\bar{q}(\downarrow)=\underbrace{\frac{1}{\sqrt{2}}\left(
\frac{t(\uparrow)\bar{q}(\downarrow)-t(\downarrow)\bar{q}(\uparrow)}{\sqrt{2}}
\right)}_{spin \,\,0}+\underbrace{\frac{1}{\sqrt{2}}\left(
\frac{t(\uparrow)\bar{q}(\downarrow)+t(\downarrow)\bar{q}(\uparrow)}{\sqrt{2}}
\right)}_{spin\,\,1}\,.
\end{equation}
The coherent superposition will become decoherent on two counts. First the
system oscillates between the two mass eigenstates with a time scale 
$t_{{\rm decoherence}}\approx1/\Delta m_{T}\approx 6\cdot10^{-22}s$ 
characterized by the mass 
difference $\Delta m_{T}=m_{T^{\ast}}-m_{T} \approx (m_{b}/m_{t})\Delta m_{B}
\approx 1 {\rm MeV}$ where $ \Delta m_{B}=m_{B^{\ast}}-m_{B}$. Loss of 
coherence through the decay 
$T^{\ast}\to T+\gamma$ can be neglected since it sets in much later at a time 
scale $t_{{\rm decay}} \approx 6 \cdot 10^{-17}s$ \cite{Grossman:2008qh}. Thus
the depolarization time scale is set by $t_{{\rm decoherence}}$ and
is larger than the tradional estimate based on 
$\Lambda^{-1}_{QCD}\cong 10^{-23}s$ by a factor of 60. Altogether, the top 
quark has decayed after $\tau_{t}=4.6\cdot 10^{-25}s$ much before 
depolarization sets in at $t_{{\rm decoherence}}\approx 6\cdot10^{-22}s$. One
concludes that the top quark retains its polarization which it has at birth 
when it decays.

The decay of polarized top quarks and the corresponding spin-momentum 
correlations in these decays will not be discussed in these lectures. A 
discussion of the spin-momentum correlations and their NLO QCD corrections
can be found in \cite{Fischer:2001gp,Fischer:1998gsa,Groote:2006kq}. We 
mention that top quarks produced at $e^{+}e^{-}$-colliders also possess a 
high degree of polarization which, in addition, can also be attuned by tuning 
the beam polarization.

The issue of whether the top quark retains its original polarization when it 
decays is also of importance in the case of hadronically produced top quark
pairs. Although the single top (or antitop) polarization is zero because
parity is conserved in the hadronic production process there are sizable 
spin-spin correlations of the top and antitop quark spins which give important 
information on the $(t \bar{t})$-production process 
(see e.g. \cite{Bernreuther:2008ju}). 

\subsection {Dominance of the decay $t \to X_{b} +W^{+} $} 

From the unitarity of the KM--matrix one has the relation
\begin{equation}
\underbrace {|V_{ub}|^{2}}_{(\approx 0.004)^{2}}\quad +\quad
\underbrace {|V_{cb}|^{2}}_{(\approx 0.04)^{2}}\quad+\quad |V_{tb}|^{2}=1\,\,. 
\end{equation}
One concludes that $ V_{tb} \approx 1$. There are a number of other SM decays 
such as $t\to X_{s}+W^{+}$ which are negligible compared to the dominant
mode $t \to X_{b} +W^{+}\,\, $\footnote{In order to simplify the notation we 
shall in the following refer
to the decay $t \to X_{b} +W^{+} $ as $t \to b +W^{+} $.}.

\subsection{Rate ratio of $t \to b+W^{+}\,(\to leptons)$ and 
$t \to b+W^{+}\,(\to hadrons )$}

Let us list the possible leptonic and hadronic decay modes of the the $W^{+}$.
For the leptonic modes one has the three modes 
\begin{equation}
W^{+} \quad \to \quad (\tau^{+}\nu_{\tau}),\,\, (\mu^{+}\nu_{\mu}),
\,\,(e^{+}\nu_{e}) \qquad \text{weight}: 3
\end{equation}
When listing the weight factor we have neglected lepton mass effects. 

For the hadronic modes one has
\begin{eqnarray}
\label{rateratio}
W^{+}\quad &\to&\quad c\bar{b},\,\,c\bar{s},\,\,c\bar{d}
\qquad  \text{weight}: 1\,\otimes 3 \,\,( colour \,summation) \nonumber \\
\quad&\to&\quad u\bar{b},\,u\bar{s},\,\,u\bar{d}
\qquad \text{weight}: 1\,\otimes 3\,\, (colour\, summation)
\end{eqnarray}
Again mass effects have been neglected.
In (\ref{rateratio}) we have summed over the respective three modes using again
the unitarity of the KM-matrix 
$\sum_{j=b,s,d}|V_{c\,j}|^{2}=1$ and $\sum_{j=b,s,d}|V_{u\,j}|^{2}=1$.
In addition one has to add in a factor of three from colour summation.
One thus obtains
\begin{equation}
\frac{\Gamma\big(t \to b+W^{+}\,(\to leptons)\big)}
{\Gamma\big(t \to b+W^{+}\,(\to hadrons )\big)}=\frac{3}{6}\,\,\,. 
\end{equation}

\subsection{ Width of the top quark}

As mentioned before the top quark decays almost 100\%\, to 
$t \to b + W^+$ in the SM. The other SM decay modes are negligible. 
Let us list the theoretical values of the SM decay width and radiative 
corrections relative to the Born term width 
(\,$\Gamma(\rm{Born})=1.56$ GeV for $m_{b}=0$).

\begin{alignat}{3}
\label{ratenum}
\frac{1}{\Gamma({\rm Born})}\,\,\Gamma_{t \to b + W^+} =  &\quad 1 &\qquad& \text{Born LO}\nonumber \\
  &- 0.27 \% &\qquad&  \text{Born} \quad m_{b}\neq 0 \nonumber \\
&- 8.5 \%  &\qquad&  \text{QCD} \quad\text{NLO} \quad  
\cite{Jezabek:1988iv} \nonumber  \\
&+1.55 \% &\qquad&  \text{electroweak \quad  NLO }
\quad \cite{Denner:1990ns,Eilam:1991iz}
\nonumber \\
&-1.56 \%&\qquad& \text{\rm finite}\,\, W^{+}-\text{width} 
\quad \nonumber \quad \cite{Jezabek:1988iv} \\
&-2.25 \% &\qquad&  \text{QCD \quad NNLO } 
\quad \cite{Blokland:2004ye,Blokland:2005vq}   
\end{alignat}
The NLO and NNLO QCD corrections and the NLO electroweak corections will be
discussed in Sec.~2. The finite width corrections will be discussed in Sec.~5.
 
It is interesting to note that the non-SM decay width into a charged Higgs
$t \to b + H^{+}$ can become comparable in size to the SM decay width
$t \to b + W^{+}$ for small and large values of $\tan\beta$ if $m_{H^{+}}$
is not too close to the phase space boundary (see e.g. 
\cite{Bernreuther:2008ju}). A precise measurement of the
top quark decay width could therefore provide stringent exclusion regions in 
the ($\tan\beta,\,m_{H^{+}}$)-parameter space of Two-Higgs-Doublet models
which contain a charged Higgs boson. 

The measurement of the top quark decay width is not simple at hadron 
colliders. In principle there are two methods to experimentally get a handle
on the decay width $\Gamma_{t}$ or lifetime $\tau_{t}=1/\Gamma_{t}$ of the
top quark. One can 
attempt to measure the mean decay length in the laboratory which is given by 
the mean decay length \footnote{We have employed a mixed notation in the
last equality of Eq.(\ref{decaylength})  where we set 
$c=1$ for the quantities in the square bracket $[p_{lab}/m]$. }
\begin{equation}
\label{decaylength}
\bar{s}=v_{lab}\cdot \tau_{lab}=\beta \gamma\,\, c\cdot \tau= 
\left[\frac{p_{lab}}{m} \right]c \cdot\tau\,\,.
\end{equation} 
where $v_{lab}=\beta c$ and $\tau_{lab}=\gamma \tau$, and
$\beta= p_{lab}c/E_{lab}$ and $\gamma=E_{lab}/(mc^{2})$.
That this measurement is 
difficult is illustrated by the 
following example. Take a top quark width of 1.43 GeV. The laboratory 
momentum of the top quark $p_{lab}$ must have the astronomically high value 
of $\approx 10^{15}$ GeV to produce a mean decay length of 1mm. Nevertheless 
CDF has attempted such a measurement using information on the magnitude of the
impact parameter of the charged lepton with respect to the collision vertex. 
CDF puts a 95\% confidence level upper limit of 
$1.8\cdot10^{-13}s$ on the lifetime of the top quark which corresponds to a 
$3.7\cdot10^{-12}$ GeV lower limit on the top quark width \cite{cdf06}. 
Naturally, this is not
a very useful bound. CDF also provides an upper limit on the top quark width
by a fit of the reconstructed top quark mass to a Breit-Wigner shape function.
The reult is $\Gamma_{t}< 13.1$ GeV at 95\% C.L. \cite{cdf08}. The upper
bound is still nine times larger than the expected SM width. 

An indirect way of determining the top quark width relies heavily on
the validity of the SM. The suggestion is to measure
the branching ratio ${\cal B}(t\to bW)= 
\Gamma(t \to bW)/\,\Gamma(t \to {\rm all})$. This could be done e.g. by 
measuring the rate of top quark pair production followed by their decays 
$t \to bW$, i.e. by measuring 
$\sigma_{t\bar{t}}\cdot({\cal B}(t\to bW))^{2}$.
Assuming that one can reliably calculate $\sigma_{t\bar{t}}$ one can then
extract ${\cal B}(t\to bW)$ (see e.g. \cite{Chakraborty:2003iw}). In the 
simplest version of this approach one takes
the SM value $\Gamma(t \to bW)$ to determine the width of the top quark through
$\Gamma(t \to {\rm all})=\Gamma(t \to bW)/{\cal B}(t\to bW)$. In a more 
sophisticated approach one uses single-top production to extract the
parameters that determine the partial width $\Gamma(t \to bW)$ 
\cite{Carlson:1995ck}.

We mention that a much improved
determination of the top quark width with an uncertainty of 
$\Delta\Gamma \approx 30$
MeV can be expected from a multi-parameter scan of the 
threshold region of $(t\bar{t})$-production at the ILC \cite{Martinez:2002st}.

\subsection{ Top quark yield}
At the LHC top quark pairs will be produced quite copiously in 
7 on 7 TeV proton-proton collisions. After a one-year probation
run at reduced energies and luminosities starting in the end of 2009 the LHC 
will start running at full energy in 2010 with a low luminosity run of 
$L \approx 10^{33}cm^{-2}s^{-1}$.
After a luminosity upgrade around the year 2017 the high luminosity run  
will have $L \approx 10^{34}cm^{-2}s^{-1}$.
Multiply these numbers with $\sigma(t\bar{t})\sim 825\, pb=825
\times10^{-36}cm^{2}$ to obtain $\approx 1\,(10) \,(t\bar{t})$-pairs every 
second for the low (high) luminosity run.

Top quark pair production at the Tevatron II (1 on 1 TeV 
$p\bar{p}$-collisions) occurs at a reduced rate. Because the energy of the 
Tevatron II
is lower, the $(t\bar{t})$-production cross section is down by a factor of 
$\approx 100$. In addition, the Tevatron II luminosity is down by 
a factor of $\approx 10$ compared to the LHC low luminosity run. Taking
these two factors into account one has $\approx 
1\times 10^{-3} \,(t\bar{t})$-pairs per second at the Tevatron II.

In the SM single top production cross section in $(p\bar{p})$- and 
$(pp)$-collisions is down by a factor of $\approx 3$
compared to top quark pair production.
The nice feature of single top production is that the top quarks are 
polarized since the production of single top quarks proceeds through
weak interactions. The polarization can be calculated to be close to
100\% (see e.g. \cite{Bernreuther:2008ju}).

At the ILC $(t\bar{t})$-production will occur at a somewhat reduced rate 
compared to the LHC. At
500 GeV the NLO rate is $\sigma(t\bar{t})\sim 0.5\, pb=0.5
\times10^{-36}cm^{2}$ (see e.g.~\cite
{Groote:1995yc,Groote:2009zk}) which gives 
$10^{-2}$ $(t\bar{t})$-events per second
assuming a luminosity of $L \approx 2\cdot 10^{34}cm^{-2}s^{-1}$.

\subsection{Polarization of $W^{+}$ gauge boson }   
The decay $t \to b + W^{+}$ is weak and therefore the $W^{+}$-boson is 
in general expected to be polarized. We shall refer to the
three partial rates that correspond to the three polarization states of the 
$W^{+}$-boson as longitudinal $(\Gamma_{L})$, transverse-plus 
$(\Gamma_{+})$ and 
transverse-minus $(\Gamma_{-})$.
At leading order (LO) the results 
for the helicity fractions ${\cal G}_{i}=\Gamma_{i}/\Gamma \quad (i=L,+,-)$
(or, in another language, for the normalized diagonal density matrix elements
of the $W^{+}$-boson $\rho_{00}$, $\rho_{++}$ and $\rho_{--}$) 
are\footnote{The 
helicities of the
$W$-boson are alternatively labelled by $(L,+,-)$, $(0,+1,-1)$ or by 
$(L,T_{+},T_{-})$.}
\begin{equation}
\label{fractions}
{\cal G}_{L}:{\cal G}_{+}:{\cal G}_{-}=\frac{1}{1+2y^{2}}:0
:\frac{2y^{2}}{1+2y^{2}}\,, 
\end{equation}
where $y^{2}=m_{W}^{2}/m_{t}^{2}= 0.211$ with $m_{b}=0$.  
Numerically one has
\begin{equation}
{\cal G}_{L}:{\cal G}_{+}:{\cal G}_{-}
=0.703:0:0.297 \,\,.
\end{equation}
Note that ${\cal G}_{L}+{\cal G}_{+}+{\cal G}_{-}=1$.
In comparison, an unpolarized $W^{+}$ would correspond to 
\begin{equation}
{\cal G}_{L}:{\cal G}_{+}:{\cal G}_{-}
=1/3:1/3:1/3 \,\,.
\end{equation}
\subsection{ Dominance of the longitudinal mode }
As $(m_{W}/m_{t}) \to 0$ the longitudinal polarization vector becomes
increasingly parallel to $q^{\mu}$ (see e.g.~\cite{Korner:2003zq}),\,{\it viz.}
\begin{equation}
\epsilon^{\mu}_{L}= \frac{1}{m_{W}}\Big(q^{\mu} + {\cal O}(m_{W}/m_{t})\Big) 
\,\,.
\end{equation}
Therefore the longitudinal mode dominates in the large top quark mass limit.
In fact, from 
$q_{\mu}\bar{u}_{b}\gamma^{\mu}(1-\gamma_{5})u_{t}
= m_{t}\bar{u}_{b}(1+\gamma_{5})u_{t}$
one concludes from dimensional arguments that 
$\Gamma_{L} \sim G_{F} m_{t}^{3}$ whereas 
$\Gamma_{\pm} \sim G_{F}m_{t}m_{W}^{2}$ or 
$\Gamma_{\pm}/\Gamma_{L}\sim m_{W}^{2}/m_{t}^{2}$.

An explicit calculation shows that $\Gamma_{+}=0$ at LO for $m_{b}=0$ 
(see Eq.(\ref{fractions})).
Looking at Fig.~1 the vanishing of the LO transverse-plus rate 
$\Gamma_{+}$ can be understood from angular momentum conservation. First
remember that a massless left-chiral fermion is left-handed as drawn in
Fig.~1. At LO one has a back-to-back decay configuration. Therefore the
$W^{+}$-boson cannot be right-handed because the $m$-quantum numbers in the 
final state would add up to $3/2$ which cannot be reached by the spin 1/2
top quark in the initial state. At NLO (or any higher order) the decay 
$t\rightarrow b+g +W^{+}$ is, in general, no longer back-to-back 
as illustrated in Fig.~1 and one anticipates that $\Gamma_{+}\neq 0$ at NLO 
and at any higher order. This is, in fact borne out by the NLO calculation
to be described later on. The physics interest lies in the fact that 
nonvanishing transverse-plus helicity rates can also be generated by non-SM 
right-chiral ($t\bar{b}$)-currents. In order to unambigously identify
non-SM contributions to the transverse-plus helicity rate it is therefore 
important to get a quantitative handle on the size of the SM higher order 
radiative correction contributions to the transverse-plus helicity rate.

\vspace{0.5cm}
\begin{figure}[h]
\begin{center}
 \includegraphics[width=10.5cm]{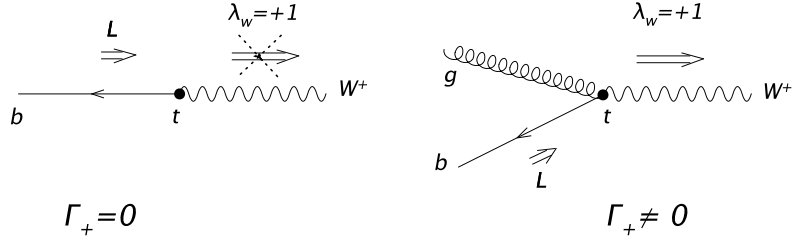}
\caption{Angular momentum conservation for $t \to b+W^{+}$ and for
$t \to b+W^{+}+g$.}
\end{center}
\end{figure}
\vspace{0.5cm}
\subsection { Measurement of the helicity fractions of the $W^{+}$ through the 
angular decay distribution in its decay  } 
The $W^{+}$ decays weakly to $(l^{+}\nu_{l})$ or to $(\bar{q}_{i}q_{j})$.
The angular decay distribution can therefore be utilized to analyze the 
polarization of the decaying $W^{+}$, i.e. the $W^{+}$ is self-analyzing.

The $W^{+}$ has the three (diagonal) polarization states $L$, $T_{+}$ and
$T_{-}$ the weights of which  
are determined by the three partial
helicity rates $\Gamma_{L}$ and $\Gamma_{\pm}$. 
As we shall explicitly derive further on, the angular decay distribution for 
$t\rightarrow b+W^{+}(\to l^{+}+\nu_{l})$ reads 
\begin{figure}[!htb]
\centering
\includegraphics[width=70mm]{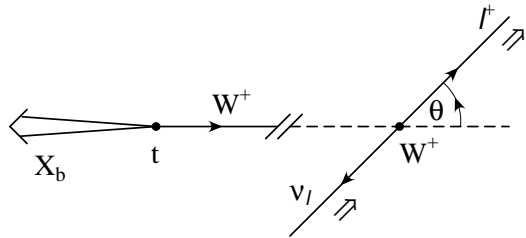}
\caption{Definition of polar angle $\theta$ in the $W^{+}$ rest system.}
\end{figure} 
\begin{equation}
\label{angdist1}
    \frac{d \Gamma}{d \cos \theta}  = 
    \frac{3}{4} \sin^2 \theta\  \Gamma_L +
    \frac{3}{8} (1 + \cos \theta)^2  \Gamma_{+} +
    \frac{3}{8} (1 - \cos \theta)^2  \Gamma_{-}\,, 
\end{equation}  
where the polar angle $\theta$ is measured in the $W$--rest frame as shown
in Fig.~2. Integrating over $ \cos \theta $ one recovers the total rate
\,\, $  \Gamma = \Gamma_{L} + \Gamma_{+} + \Gamma_{-}$.
If the $W^{+}$ were unpolarized one would have $\Gamma_{L} = \Gamma_{+} 
= \Gamma_{-}=\Gamma/3$ resulting
in a flat decay distribution $d\Gamma/d\!\cos\theta=\Gamma/2$\,.

One can also define a forward-backward asymmetry by considering the
rate in the forward hemisphere $\Gamma_{F}$ and in the backward hemisphere 
$\Gamma_{B}$ in the $W^{+}$-rest frame. The forward-backward asymmetry 
$A_{FB}$ is then given by
\begin{equation}
A_{FB}=\frac{\Gamma_{F}-\Gamma_{B}}{\Gamma_{F}+\Gamma_{B}}=
\frac{3}{4}\,\, \frac{\Gamma_{+}-\Gamma_{-}}{\Gamma_{L}+\Gamma_{+}+\Gamma_{-}}
\,\,.
\end{equation}
At the Born term level one has
\begin{equation}
A_{FB}({\rm Born})=-\frac{3}{4}\,\,\frac{2y^{2}}{1+2y^{2}}=-0.22\,\,.
\end{equation}
The forward-backward asymmetry is negative, i.e. one has more leptons
in the backward hemisphere than in the forward hemisphere. The numerical
value of the forward-backward asymmetry is not very large on account of the 
dominance of the longitudinal mode.

It is always useful to check on the correctness of the sign of the parity 
violating term
proportional to ($\pm\cos\theta$) and thereby on the sign of $A_{FB}$.
This is again easily done by considering the collinear cases 
$\cos \theta=\pm 1$ and appealing to angular momentum conservation. And,
in fact, Eq.(\ref{angdist1}) shows that the mode $\Gamma_{-}$ decouples
in the forward direction $\cos \theta=1$
(and vice versa $\Gamma_{+}$ decouples in the backward direction) as can be
appreciated from the helicity configurations in Fig.~2. This 
implies that $\Gamma_{+}$ favours forward leptons $\ell^{+}$ leading to
energetic leptons in the $t$-rest frame whereas
$\Gamma_{-}$ favours backward $\ell^{+}$ leading to less energetic leptons in
the $t$-rest frame. As we have seen $\Gamma_{+}=0$ at LO so that one expects 
a softer lepton spectrum in the $t$-rest frame then in the case of the decay 
of an unpolarized $W^{+}$.

\section{Top quark decay rate}

\subsection{Leading (LO) rate}
We shall calculate the leading order rate in three different ways for
pedagogical reasons. The first way is the traditional covariant way
where no particular sophistication is needed. In the second way we use the 
helicity methods which has the advantage that by calculating the helicity
amplitudes one has the full spin information of the decay at hand. In the
third method we use the optical theorem which serves the purpose of 
introducing rather sophisticated technical material in a simple setting
which are needed later on in the higher order calculations.
 
\subsubsection{Covariant method}
The matrix element for the decay $t \to b + W^{+}$ $(p_{t}=p_{b}+q)$ is given 
by
\begin{equation}
M=-i\frac{g_{w}}{2\sqrt{2}} V_{tb}\bar{u}_{b}
\gamma^{\mu}(1-\gamma_{5})u_{t}\eps^{*}_{\mu}\,\,. 
\end{equation}
Upon squaring and summing over the spins one obtains
\begin{equation}
|\overline{M}|^{2}= \sum_{spins}\frac{g_{w}^{2}}{8}|V_{tb}|^{2}
\Big(\bar{u}_{b}
\gamma^{\mu}(1-\gamma_{5})u_{t}\eps^{*}_{\mu}\Big) \Big(\bar{u}_{b}
\gamma^{\nu}(1-\gamma_{5})u_{t}\eps^{*}_{\nu}\Big)^{\dagger}\,,
\end{equation}
where we write $\sum_{spins}|M|^{2}=|\overline{M}|^{2}$.
Use of the completeness relations
\begin{equation}
\label{completenessfermion}
\sum_{\pm 1/2}u \bar{u}=\slashed{p}+m
\end{equation}
and
\begin{equation}
\label{completenessboson}
 \sum_{0,\pm}\epsilon^{\mu }(m)\epsilon^{\ast \nu }(m) = 
-g^{\mu \nu}+\frac{q^{\mu}q^{\nu}}{m_{W}^{2}} 
\end{equation}
leads to $(m_{b}=0)$
\begin{align}
\label{msquared1}
|\overline{M}|^{2}=&\frac{g_{w}^{2}}{8}|V_{tb}|^{2}{\rm Tr}
\left\{\slashed{p}_{b}\gamma^{\mu}(1-\gamma_{5})
(\slashed{p}_{t}+m_{t})\gamma^{\nu}(1-\gamma_{5})\right\} \Big(-g_{\mu\nu}+\frac{q_{\mu}q_{\nu}}
{m_{W}^{2}} \Big)
\nonumber \\
=&\frac{g_{\omega}^{2}}{8}|V_{tb}|^{2}\,2{\rm Tr} 
\left\{\slashed{p}_{b}\gamma^{\mu}
\slashed{p}_{t}\gamma^{\nu}\right\} \Big(-g_{\mu\nu}+\frac{q_{\mu}q_{\nu}}
{m_{W}^{2}} \Big)
\nonumber \\
=&\frac{g_{w}^{2}}{8}|V_{tb}|^{2}\,8\Big((p_{t}p_{b})+
\frac{1}{m_{W}^{2}}2(p_{b}p_{t})\,(p_{t}q)\Big)\,\,. 
\end{align}

Using four-momentum conservation $p_{t}=p_{b}+q$ and the mass shell conditions
$p_{b}^{2}=m_{b}^{2}=0$ and $q^{2}=m_{W}^{2}$ one obtains
\begin{equation}
|\overline{M}|^{2}=\frac{g_{w}^{2}}{8}|V_{tb}|^{2}\,4 m_{t}^{2}
\frac{1-y^{2}}{y^{2}}(1+2y^{2})\,\,.
\nonumber
\end{equation}
The rate can be computed using the two--body decay formula 
\begin{equation}
\label{rate}
\Gamma=\frac{1}{2s_{t}+1} R_{2}\,\big[|\overline{M}|^{2}\big]\,,
\end{equation}
where $R_{2}$ denotes 
the two-body phase space integral \cite{Peskin:1995ev}. We symbolically
write $R_{2}\,\big[|\overline{M}|^{2}\big]$ for the two-body phase space
integration over the squared matrix element $|\overline{M}|^{2}$, i.e.
we write  
\begin{equation}
\label{msquared2}
R_{2}\,\big[\,|\overline{M}|^{2}\big]=\frac{1}{2m_{t}}
\int \frac{1}{(2\pi)^{3}}\frac{d^{3}q}{2\,E_{W}}
\int \frac{1}{(2\pi)^{3}}\frac{d^{3}p_{b}}{2\,E_{b}}\,
(2\pi)^{4}\,\delta^{(4)}(p_{t}-p_{b}-q)\,\,|\overline{M}|^{2}\,\,. 
\end{equation}
In order to stay general we calculate $R_{2}$ for $m_{b}\neq 0$. The phase
space integral will be evaluated in the top quark rest system. We write 
$1/(2E_{W})=\int dE_{W}\, \delta(q^{2}-m_{W}^{2})
=\int dE_{W}\, \delta(E_{W}^{2}- |\vec{q}|^{2}-m_{W}^{2})$, where we implicitly
take the positive energy solution $E_{W}=+\sqrt{|\vec{q}|^{2}+m_{W}^{2}}$.
The corresponding relation for the bottom quark energy reads
$1/(2E_{b})=\int dE_{b}\, \delta(p_{b}^{2}-m_{b}^{2})$. Using these two
relations one converts the three-dimensional integrations in 
(\ref{msquared2}) into four-dimensional integrations. One obtains 
\begin{equation}
R_{2}\,\big[\,|\overline{M}|^{2}\big]=\frac{1}{8\pi^{2}m_{t}}\int d^{4}q \int d^{4}p_{b}\,
\delta(q^{2}-m_{W}^{2})\,\delta(p_{b}^{2}-m_{b}^{2})\,
\delta^{(4)}(p_{t}-p_{b}-q)\,\,|\overline{M}|^{2}  \,\,.
\end{equation}
The integration over $d^{4}p_{b}$ can be done with the result that the argument
of the second $\delta$-function becomes $(p_{t}-q)^{2}-m_{b}^{2}=
m_{t}^{2}-2m_{t}E_{W}+m_{W}^{2}-m_{b}^{2}$, i.e.
\begin{equation}
\label{phasespace}
R_{2}\,\big[\,|\overline{M}|^{2}\big]=\frac{1}{8\pi^{2}m_{t}}\int d^{4}q \, 
\delta(q^{2}-m_{W}^{2})\,\delta\big((p_{t}-q)^{2}-m_{b}^{2}\big)\,\,|\overline{M}|^{2} \,\,.  
\end{equation}
Next one integrates over
$dE_{W}=d(2m_{t}E_{W})/2m_{t}$ with the result that the argument of the
remaining $\delta$-function becomes $q^{2}-m_{W}^{2}
=E_{W}^{2}-|\vec{q}|^{2}-m_{W}^{2} \to (m_{t}^{2}+m_{W}^{2}-m_{b}^{2})^{2}
/(4m_{t}^{2})-|\vec{q}|^{2}-m_{W}^{2}$. The remaining integration over
$d^{3}q$ can be done using spherical coordinates such that
$d^{3}q \to d\Omega |\vec{q}|^{2}d|\vec{q}|^{2}=\frac{1}{2}|\vec{q}|
d|\vec{q}|^{2}$. The result is
\begin{equation}
R_{2}\,\big[\,|\overline{M}|^{2}\big]=\frac{1}{8\pi}\frac{1}{m_{t}^{2}}\,|\vec{q}|\,\,|\overline{M}|^{2}\,\,,
\end{equation}
where $|\vec{q}|= \sqrt{\lambda(m_{t}^{2},m_{W}^{2},m_{b}^{2})}/(2m_{t})$ 
is the magnitude of the momentum of
the $W^{+}$-boson in the top quark rest ($t$--rest) frame and
where $\lambda(a,b,c)=(a^{2}+b^{2}+c^{2}-2ab-2ac-2bc)$ is K\"all\'en's 
function. Naturally we could have calculated the two-body phase space
$R_{2}$ directly without including the squared matrix element 
$|\overline{M}|^{2}$ in the integrand as long as the kinematic variables in 
$|\overline{M}|^{2}$ are fixed according to four-momentum conservation and the
 mass-shell conditions.

We now return to the approximation $m_{b}=0$ where 
$|\vec{q}\,|= m_{t}(1-y^{2})/2$. Substituting the matrix element squared
(\ref{msquared1}) into the rate formula (\ref{rate}) one obtains 

\begin{equation}
\Gamma({\rm Born})=\Gamma_{0}\,(1-y^{2})^{2}(1+2y^{2})\,,
\end{equation}
where $\Gamma_{0}$ is the $m_{W}=0$ Born term rate ($g_{\omega}^{2}/(8m_{W}^{2})=G_{F}/\sqrt{2}$)
\begin{equation}
\Gamma_{0}=\frac{G_{F} m_{t}^{3}}{8\pi\sqrt{2}}|V_{tb}|^{2}\,. 
\end{equation}
\subsubsection{Helicity amplitude method}
The helicity amplitudes for $t \to b +W^{+}$ can be calculated from the
transition matrix element by using spinors and polarization vectors with
definite helicities $\lambda_{t},\,\lambda_{b}$ and $\lambda_{W}$. One needs to 
calculate (we omit the coupling factor 
$-i\frac{g_{w}}{2\sqrt{2}} V_{tb}$)
\begin{equation}
\qquad\qquad H_{\lambda_{t};\lambda_{b}\lambda_{W}}=\bar{u}_{b}(\lambda_{b})
\gamma^{\mu}(1-\gamma_{5})u_{t}(\lambda_{t})\eps^{*}_{\mu}(\lambda_{W})\,\,.
\end{equation}
We shall work in the $t$--rest system with the $z$-axis along the 
$W^{+}$ (see Fig.~3) such that $\lambda_{t}=-\lambda_{b}+\lambda_{W}$.
In order to be general we keep $m_{b}\neq 0$.
\begin{figure}[!htb]
\begin{center}
\includegraphics[width=80mm]{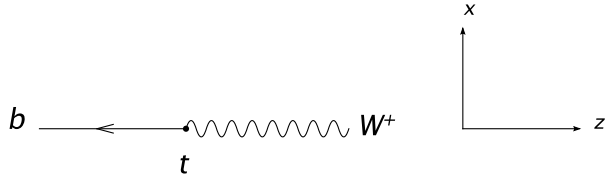}
\caption{Definition of the the two--body coordinate system in the top
quark rest system.}
\end{center}
\end{figure} 

Let us collect the relevant $t$--rest system spinor and polarization 
vector expressions. For the helicity spinors one has
\begin{align}
u_{t}(1/2)=&\sqrt{2m_{t}} { \chi_{+} \choose 0}\,,
\qquad  u_{b}(1/2)=\sqrt{E_{b}+m_{b}}\left( 
\begin{array}{c} 
 \chi_{-} \\
 \frac{|\vec{q}|}{E_{b}+m_{b}}\chi_{-}
 \end{array}
\right)\,, \nonumber \\
u_{t}(-1/2)=&\sqrt{2m_{t}} { \chi_{-} \choose 0}\,,
\qquad u_{b}(-1/2)=\sqrt{E_{b}+m_{b}}\left( 
\begin{array}{c} 
-\chi_{+}\\
\frac{|\vec{q}|}{E_{b}+m_{b}}\chi_{+}
 \end{array}
\right) \,,
\end{align}
where $\chi_{\pm}$ are Pauli spinors given by 
$\chi_{+}= {1 \choose 0}$ and $\chi_{-}= {0 \choose 1}$\,.

The helicity polarization four-vectors of the $W^{+}$ read
\begin{align}
\label{polvec1}
\eps_{\mu}^{*}(\pm1)&=\frac{1}{\sqrt{2}}\left(0;\pm1,-i,0\right)\,,
\nonumber \\
\eps_{\mu}^{*}(0)&=\frac{1}{\sqrt{q^{2}}}\left(|\vec{q}|;0,0,-q_{0}\right)\,.
\end{align}

There are altogether four possible helicity configurations in $t\to b +W^{+}$
which are listed in
Table~1.
\begin{table}[h]
\begin{center}
\begin{tabular}{ccc}
\hline\hline
$\lambda_{t}$ \,\,&$\lambda_{b}\,\,$ &$\lambda_{W}$ \\ 
\hline\hline
1/2 &-1/2 & 0 \\
-1/2 & 1/2 &0 \\
1/2 &1/2 &1 \\
-1/2 &-1/2&-1 \\ \hline
\end{tabular} 
\end{center}
\caption{\label{tab:helconfig} 
Helicity configurations in $t \to b + W^{+}$. 
}
\end{table}

For the helicity amplitudes $H_{\lambda_{t};\lambda_{b}\lambda_{W}}$
\, ($Q_{\pm}=(m_{t}\pm m_{b})^{2}-q^{2}$) one obtains
\begin{alignat}{2}
\label{helamp}
\sqrt{q^{2}}H_{\frac{1}{2};-\frac{1}{2}0}&=-m_{t}(\sqrt{Q_{+}}+\sqrt{Q_{-}})
+m_{b}(\sqrt{Q_{+}}-\sqrt{Q_{-}}) \quad &\stackrel{m_{b}\to 0}{=} 
\quad &-2m_{t}^{2}\sqrt{1-y^{2}}\,,
\nonumber \\
\sqrt{q^{2}}H_{-\frac{1}{2};\frac{1}{2}0}&=-m_{t}(\sqrt{Q_{+}}-\sqrt{Q_{-}})
+m_{b}(\sqrt{Q_{+}}+\sqrt{Q_{-}}) \quad &\stackrel{m_{b}\to 0}{=} 
\quad &\qquad 0\,,
\nonumber \\
H_{\frac{1}{2};\frac{1}{2}1}&=-\sqrt{2}(\sqrt{Q_{+}}-\sqrt{Q_{-}}) 
&\stackrel{m_{b}\to 0}{=} \quad
&\qquad 0 \,,\nonumber \\
H_{-\frac{1}{2};-\frac{1}{2}-1}&=-\sqrt{2}(\sqrt{Q_{+}}+\sqrt{Q_{-}}) 
&\stackrel{m_{b}\to 0}{=} 
\quad& -2\sqrt{2}m_{t}\sqrt{1-y^{2}}\,,
\end{alignat}
where we have included both the $m_{b} \neq 0$ and $m_{b}=0$ results in
(\ref{helamp}).
The squared matrix element $|\overline{M}|^{2}$ finally is given by
\begin{eqnarray}
\label{msquared3}
|\overline{M}|^{2}=\sum_{\lambda_{t}=-\lambda_{b}+\lambda_{W}}
|H_{\lambda_{t};\lambda_{b}\lambda_{W}}|^{2}&=&
\underbrace{|H_{\frac{1}{2};-\frac{1}{2}0}|^{2}+|H_{-\frac{1}{2};\frac{1}{2}0}|^{2}}_{L}
+\underbrace{|H_{\frac{1}{2};\frac{1}{2}1}|^{2}}_{T_{+}}
+\underbrace{|H_{-\frac{1}{2};-\frac{1}{2}-1}|^{2}}_{T_{-}}
\nonumber \\
&=&4m_{t}^{2}\frac{(1-y^{2})}{y^{2}}
\Big(\underbrace{\phantom{a}1\phantom{a}}_L\quad+ \underbrace{0}_{T_{+}}
\quad+\underbrace{2y^{2}}_{T_{-}}\Big)\,,
\end{eqnarray}
where we have set $m_{b}=0$ in the second line of (\ref{msquared3}).
The result agrees with the covariant calculation (see Eq.(\ref{msquared1})). 
The advantage of the 
helicity method
is that one can separately identify the three (diagonal) helicity contributions
of the $W^{+}$ boson $L$, $T_{+}$ and $T_{-}$ as indicated in 
Eq.(\ref{msquared3}). In fact, the helicity amplitudes contain the complete
spin information of the process. Thus one can easily calculate other
polarization effects using the helicity amplitudes such as the decay of 
polarized top quarks, the polarization of the bottom quark and 
polarization correlation effects. 
$m_{b} \neq 0$ effects are 
easily included by using the $m_{b} \neq 0$ helicity amplitudes in 
Eq.(\ref{helamp}). One can also define covariant helicity projectors which 
allow one to directly calculate the longitudinal, the transverse-plus
and transverse-minus helicity rates without taking recourse to the helicity 
amplitudes. This will be described in Sec.~5. 
\subsubsection{Optical theorem method and cutting rules}
In this subsection we shall use yet another method to calculate the leading
order rate for $t \to b +W^{+}$ using the optical theorem. Whereas the
optical theorem method does not offer particular technical advantages in LO 
calculations it is the method of choice for higher order calculations as e.g. 
the calculation of the NNLO rate to be described later on. The reason is 
simply that the phase space integrations in the NNLO radiative correction
calculations become prohibitively complicated and cannot be automated as 
easily as higher order loop calculations. We present the optical theorem method
for the LO case for pedagogical reasons because the LO discussion allows us
to introduce concepts which are also needed in the NNLO radiative correction
calculation to be described later on.
\begin{figure}[!h]
\begin{center}
  \includegraphics[height=24mm]{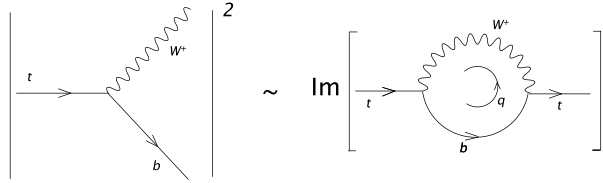}
\caption{Illustration of the optical theorem method to calculate the LO
top quark width.}
\end{center}
\end{figure}

The optical theorem relates the width $\Gamma$ of a particle to the imaginary 
part of the self-energy contribution $\Sigma$ of the particle. In the top 
quark case one has \footnote{A very nice discussion of the optical theorem
and related technical material relevant to top quark decays can be found in 
the thesis of I.R. Blokland~\cite{Blokland:2004nd}.}
\begin{equation}
\Gamma =\frac{1}{2s_{t}+1}\frac{{\rm Im}\, \Sigma}{m_{t}}\,, 
\end{equation} 
where, for the present purposes, $\Sigma$ is the one-loop self-energy of the 
top quark as illustrated in Fig.4.

Using standard Feynman rules \cite{Peskin:1995ev,Blokland:2004nd} the 
one-loop self-energy contribution is given by $(p_{t}=p_{b}-q)$
\begin{eqnarray}
\label{moneloop}
i\Sigma_{one-loop} &=&\sum_{s_{t}}
\bar{u}(p_t , s_t)\int \frac{d^{4}q}{(2\pi)^{4}}\left(i\frac{g_{w}}{\sqrt{2}}
\gamma^\mu\frac{1-\gamma_5}{2}V_{tb}\right) \frac{i\slashed{p}_{b}}
{p_{b}^{2}+i\epsilon}\times \nonumber \\[2mm]
&& \qquad\qquad\times \left(i\frac{g_{w}}{\sqrt{2}}
\gamma^\nu\frac{1-\gamma_5}{2}V_{tb}*\right)
\frac{-i\left(g_{\mu\nu}-q_{\mu}q_{\nu}/m_{W}^{2}\right)}{q^{2}-m_{W}^{2}+i\epsilon} \, 
u(p_t, s_t)\,.
\end{eqnarray}
One can again use the completeness relation $\sum_{\pm1/2}u_{t}\bar{u}_{t}=
(\slashed{p}_{t}+m_{t})$ to rewrite Eq.(\ref{moneloop}) as a trace. 
The trace can be taken
as in Eq.(\ref{msquared1}) except that one now cannot avail of the mass-shell 
conditions $q^{2}=m_{W}^{2}$ and $p_{b}^{2}=0$. One obtains
\begin{equation} 
\Sigma_{one-loop}=\frac{1}{i}\frac{g_{w}^{2}}{8}|V_{tb}|^{2}\,
\frac{8}{m_{W}^{2}}
\int \frac{d^{4}q}{(2\pi)^{4}}
\frac{p_{t}q(2p_{t}q+2q^{2}+m_{W}^{2})+m_{t}^{2}m_{W}^{2}}
     {\big[(p_{t}+q)^{2}+i\epsilon\big]
\big[q^{2}-m_{W}^{2}+i\epsilon \big]}\,.
\end{equation}
The usual procedure is to expand the $q$-dependent numerator factors in terms 
of the $q$-dependent denominator factors $D_{q}=q^{2}-m_{W}^{2}$ and
$D_{b}=(p_{t}+q)^{2}$ in order to obtain $q$-independent
numerator factors (the corresponding integrals are called scalar integrals) 
after cancellation. We therefore write
\begin{eqnarray}
p_{t}q&=&-\frac{1}{2}(m_{t}^{2}+m_{W}^{2}+D_{q}-D_{b})\,, \nonumber \\
q^{2}&=&m_{W}^{2}+D_{q}\,\,. 
\end{eqnarray}
The contributions proportional to $D_{q}$ and $D_{b}$ cancel against the 
denominator pole factors and
their contributions can be dropped when taking the imaginary part 
since single or zero pole contributions have no imaginary part 
(see e.g.~\cite{Blokland:2004nd}). We 
therefore have
\begin{equation}
\Sigma_{one-loop}=\frac{1}{i}\frac{g_{w}^{2}}{8}|V_{tb}|^{2}\,
\frac{4m_{t}^{4}}{m_{W}^{2}}(1-y^{2})(1+2y^{2})
\int \frac{d^{4}q}{(2\pi)^{4}}
\frac{1}
 {\big[(p_{t}+q)^{2}+i\epsilon\big]\big[q^{2}-m_{W}^{2}+i\epsilon)\big]}\,\,.
\end{equation}

According to the cutting rules the discontinuity of a Feynman graph is
obtained by the product of the discontinuities of the pole factors which
are being cut, where the discontinuity of a single pole is given by 
\cite{Peskin:1995ev,Blokland:2004nd} 
\begin{equation}
{\rm Disc} \, \frac{1}{p^{2}-m^{2}+i\epsilon}=-2\pi i\,\delta(p^{2}-m^{2})\,.
\end{equation}
Furthermore, the imaginary part and the discontinuity of a graph $M$ are 
related by $2i\, {\rm Im}M={\rm Disc}\,M$. 
One therefore has
\begin{equation}
\label{imresult}
 {\rm Im}\Sigma_{one-loop}=\frac{1}{i}\frac{1}{2i} \frac{g_{w}^{2}}{8}
|V_{tb}|^{2}\,\frac{4m_{t}^{4}}{m_{W}^{2}}(1-y^{2})(1+2y^{2})
\int \frac{d^{4}q}{(2\pi)^{4}}(-2\pi i)^{2} \delta\big(
(p_{t}+q)^{2}\big)\delta(q^{2}
-m_{W}^{2})\,\,.
\end{equation}
In order to exhibit the similarity to the integral (\ref{phasespace}) we
change the integration variable $q \to -q$. One can then use the result
of Sec.~2.1.1
\begin{equation}
\int d^{4}q \, 
\delta(q^{2}-m_{W}^{2})\,\delta\big((p_{t}-q)^{2}\big)=\frac{\pi}{2}(1-y^{2)}) 
\nonumber
\end{equation}
to arrive at
\begin{equation}
\Gamma(Born)=\Gamma_{0}(1-y^{2})^{2}(1+2y^{2})\,,
\end{equation}
where, as before,
\begin{equation}
\Gamma_{0}=\frac{G_{F}\, m_{t}^{3}}{8\sqrt{2}\pi}|V_{tb}|^{2}\quad {\rm and}
\quad y=\frac{m_{W}}{m_{t}}\,.
\end{equation}
As it must be the result agrees with the covariant and helicity amplitude 
calculations.  It is
quite reassuring that the decay rate turns out to be positive definite in the 
end, as it must be, considering all the minus signs and the factors of (i)
appearing in the rate calculation using the optical theorem method.


\subsubsection{Expansion by regions and the $(m_{W}/m_{t})$-expansion}

In Sec~2.1.3 we have calculated the leading order rate by using the optical 
theorem and cutting rules to determine the imaginary part of the one-loop
self energy diagram. In this subsection we shall go one step further and
calculate the leading order rate using a $(m_{W}/m_{t})$-expansion which 
allows us to 
introduce the concepts of expansion by regions and integration-by-parts
identities. All latter three concepts are essential in the calculation of the 
NLLO rate presented in \cite{Blokland:2004ye,Blokland:2005vq}. As emphasized 
before we shall pattern the LO rate calculation after the
NNLO calculation entirely for pedagogical reasons. In the LO case the 
follow-up calculations
are simple enough to be presented in a few simple lines, whereas they are more
involved in the full NNLO calculation.

Let us summarize the main ideas of the NNLO rate calculation presented in  
\cite{Blokland:2004ye,Blokland:2005vq} which we down-size to the present LO 
case.
\begin{itemize}
\item Reduce the two-mass-scale problem $(m_{t},m_{W})$ to a one-mass-scale
problem $(m_{t})$ by expanding in the ratio $m_{W}/m_{t}$. Obtain the results 
as an expansion in powers of 
$m_{W}/m_{t}$.
\item Use dimensional regularization to regularize the UV and IR/M 
singularities
\item  Use the method of expansion by regions to calculate the 
one-loop integral~\cite{Smirnov:1994tg,Smirnov:1996ng,Beneke:1997zp}.
\end{itemize}
One has to consider the two 
regions~\cite{Smirnov:1994tg,Smirnov:1996ng,Beneke:1997zp}:
\begin{itemize} 
\item Hard region \\
The loop momentum is hard and is of ${\cal O}(m_{t})$. 
One can then expand the $W$-propagator
as a power series in $m_{W}^{2}/m_{t}^{2}\ll 1$:
\begin{equation}
\label{expan1}
\frac{1}{q^{2}-m_{W}^{2}}=\frac{1}{q^{2}}+ \frac{m_{W}^{2}}{q^{4}} +
\frac{m_{W}^{4}}{q^{6}}
+...= \frac{1}{q^{2}}\sum_{n=0}^{\infty}\left(\frac{m_{W}^{2}}{q^{2}}\right)^{n}\,\,.
\end{equation}
The massive propagator has thereby been converted into a sum of massless 
propagators.
\item Soft region \\
The momentum q flowing through the $W$ is soft. One therefore cannot use the
above expansion (\ref{expan1}) of the $W$ propagator. However, in the soft 
region one can
expand the $b$-quark propagator, {\it cif.}
\begin{equation}
\frac{1}{(p_{t}+q)^{2}}
=\frac{1}{p_{t}^{2}}
\sum_{n=0}^{\infty}\left(-\frac{2p_{t}\cdot q + q^{2}}{p_{t}^{2}} 
\right)^{n} \,\,.
\end{equation}
There is only one denominator factor in the loop integral and its imaginary 
part vanishes
\begin{equation}
{\rm Im }\int\frac{1}{i}\frac{d^{4}q}{(2\pi)^{4}}\frac{1}{(q^{2}-m_{W}^{2})} =0\,.
\end{equation}
Therefore there is no contribution from the soft region in the one-loop case.
This is different at NLO and NNLO.
\end{itemize}

What remains to be done is to evaluate integrals of the form
\begin{equation}
\label{int1}
{\rm Im}\int\frac{1}{i} \frac{d^{4}q}{(2\pi)^{4}}\frac{1}{(q^{2})^{n+1}(p_{t}+q)^{2}} 
\end{equation}
which result from the $m_{W}^{2}/q^{2}$ expansion in the hard region.
The integrals can all be reduced to one master integral by using
integration-by-parts identities.

The first term in the expansion (\ref{expan1}), $n=0$,
leads to a two--point one--loop integral of the form
\begin{equation}
\label{int2}
{\rm Im}\int\frac{1}{i} \frac{d^{4}q}{(2\pi)^{4}} \frac{1}{q^{2}(p_t+q)^{2}}\,\,.
\end{equation}
We calculate the one-loop integral directly in dimensional regularization
$(D=4-2\epsilon)$ and take its imaginary part at the end
without resorting to the cutting rules. The details of how to evaluate 
one-loop integrals in dimensional
regularization can be found in \cite{Peskin:1995ev}. One first introduces a 
one parameter Feynman parametrization, collects terms and performs a shift in 
the integration variable $(q + xp_{t}) \to q$, i.e.  
\begin{eqnarray}
\label{im1}
\int\D{q}\frac{1}{q^{2}(p_{t}+q)^{2}}&=&\int \D{q}\int_{0}^{1} dx 
  \frac{1}{\left[ (q+p_t)^{2}x+q^{2}(1-x) \right]^{2}} \nonumber\\
&=& \int \D{q}\int_{0}^{1} dx 
  \frac{1}{\left[ (q+xp_t)^{2}+p_t^{2}x(1-x) \right]^{2}}\nonumber \\
&=& \int \D{q}\int_{0}^{1} dx 
  \frac{1}{\left[ q^{2}+p_t^{2}x(1-x) \right]^{2}}\,\,.  
\end{eqnarray}
Next we do a Wick rotation $q_{0} \to iq_{0E}$.
The factor of $i$ from the Wick rotation cancels the factor of $i$ in 
the denominator of (\ref{int1}). One then does a
D-dimensional Euclidean integration over the loop 
momentum $q$, and, finally, one integrates over the Feynman parameter $x$
which results in Euler's Beta function $B(1-\epsilon,1-\epsilon)$. The sequence
of steps is represented in the following sequence of equations:
\begin{eqnarray}
\label{im2}
\im \int\frac{1}{i}\D{q}\frac{1}{q^{2}(p_{t}+q)^{2}}&=&\im \int_{0}^{1} dx 
\frac{1}{(4\pi)^{D/2}}\frac{\Gamma(2-\frac{D}{2})}{\Gamma(2)}
   \left(\frac{1}{-p_t^{2}x(1-x) }\right)^{2-\frac{D}{2}} \nonumber\\
&=& \im\frac{\Gamma(1+\epsilon)}{(4\pi)^{D/2}}
    \frac{\Gamma(\epsilon)}{\Gamma(1+\epsilon)}(-p_t^{2})^{-\epsilon}
    \int_{0}^{1} dx \, x^{-\epsilon}(1-x)^{-\epsilon} \nonumber \\
&=& \frac{\Gamma(1+\epsilon)}{(4\pi)^{D/2}}\frac{\Gamma(\epsilon)}{\Gamma(1+\epsilon)}B(1-\epsilon,1-\epsilon)
     \im(-p_t^{2})^{-\epsilon} \nonumber \\
&=& \frac{\Gamma(1+\epsilon)}{(4\pi)^{D/2}}\frac{\Gamma(\epsilon)}{\Gamma(1+\epsilon)}
     \frac{\Gamma^{2}(1-\epsilon)}{\Gamma(2-2\epsilon)}
(m_{t}^{2})^{-\eps}\sin \pi \eps\,\,.
\end{eqnarray}
We retain only the finite term in the last line of (\ref{im2}). One obtains 
\begin{equation}
\im \int \frac{1}{i}\frac{d^{4}q}{(2\pi)^{4}} \frac{1}{q^{2}(p_t+q)^{2}}=
\frac{1}{16\pi^{2}}\,\,\pi\,.
\end{equation}
We have used 
\begin{equation}
(\frac{-p_{t}^{2}}{m_{t}^{2}})^{-\eps}=e^{\ln(-p_{t}^{2}/m_{t}^{2})^{-\eps}}
=e^{-\eps\ln(-p_{t}^{2}/m_{t}^{2})}
=1-\eps\ln(p^{2}/m_{t}^{2}) +... 
\end{equation}
which leads to
\begin{equation}
\im (\frac{-p_{t}^{2}}{m_{t}^{2}})^{-\eps}=\im (-1+i0)^{-\eps}
=\sin \pi \eps=\pi\epsilon+...\,\,\,.
\end{equation} 

In addition to the integral (\ref{int2}) with $n=0$ the imaginary part of 
which we have 
just calculated we also need the imaginary parts of the integrals (\ref{int1})
with $n \ge 1$.
They can be obtained from the ``master integral'' (\ref{int2})
by integration-by-parts (IBP) techniques 
\cite{Tkachov:1981wb,Chetyrkin:1981qh}. The general procedure of reducing
a set of integrals to a set of simpler integrals is called ``reduction to 
master integrals''. In the present case this reduction is quite trivial
but can become quite involved in more general settings. The reduction 
procedure has been automated by the Laporta 
algorithm~\cite{Laporta:1996mq,Laporta:2001dd}.

{\bf Technical aside: Integration-by-parts (IBP) identities}~\cite
{Tkachov:1981wb,Chetyrkin:1981qh}.\\
In order to calculate the integral corresponding to the second term in the
expansion (\ref{expan1}) we consider the differential form 
($\partial_{\mu}:=\partial/\partial q^{\mu}$) 
\begin{equation}
\label{ibp1}
\partial_{\mu}\frac{(p_{t}+q)^{\mu}}{q^{2}(p_{t}+q)^{2}}=
\frac{\partial_{\mu}(p_{t}+q)^{\mu}}{q^{2}(p_{t}+q)^{2}}
+\frac{(p_{t}+q)^{\mu}}{q^{2}}\partial_{\mu}\frac{1}{(p_{t}+q)^{2}}
+\frac{(p_{t}+q)^{\mu}}{(p_{t}+q)^{2}}\partial_{\mu}\frac{1}{q^{2}}\,. 
\end{equation}
Differentiate carefully, i.e.
$\partial_{\mu}q^{\mu}:=\frac{\partial q^{\mu}}{\partial q^{\mu}}=
\,D=4-2\eps$, and drop the 
``surface term'' on the left-hand side. Also use 
$2p_{t}q=-m_{t}^{2}-q^{2}+(p_{t}+q)^{2}$. This gives
\begin{equation}
\label{ibp2}
\frac{1}{q^{4}(p_{t}+q)^{2}}=\frac{1}{m_{t}^{2}}
\left(\frac{2\eps-1}{q^{2}(p_{t}+q)^{2}}+\frac{1}{q^{4}}\right) \,.
\end{equation}
In dimensional regularization massless tadpole (single pole) diagrams are 
zero, i.e. one can drop the second term on the r.h.s. of (\ref{ibp2}) after 
dimensional 
integration.
 At the relevant order of $\epsilon$ one therefore has 
\begin{equation}
\im \int\frac{1}{i} \frac{d^{D}q}{(4\pi)^{D}}\frac{1}{q^{4}(p_{t}+q)^{2}}=
-\frac{1}{m_{t}^{2}}\im \int\frac{1}{i} \frac{d^{D}q}{(4\pi)^{D}}\frac{1}
{q^{2}(p_{t}+q)^{2}} \,.
\end{equation}
Going through the same exercise for $\partial_{\mu}\frac{(p_{t}+q)^{\mu}}
{(q^2)^{n+1}(p_{t}+q)^{2}}$ for $n \ge 2$ one finds
\begin{equation}
\im \int\frac{1}{i} \frac{d^{D}q}{(4\pi)^{D}}\frac{1}{(q^{2})^{n+1}
(p_{t}+q)^{2}}=0\,. 
\qquad {\rm for}\quad n \ge 2 
\end{equation}

Because the higher order terms vanish
we only need to sum the first two terms in the expansion (\ref{expan1}).
The result  
\begin{equation}
\im \int\frac{1}{i} \frac{d^{D}q}{(4\pi)^{D}}\frac{1}{(p_{t}+q)^{2}}\frac{1}{q^{2}}
\sum_{n=0}^{\infty}\left(\frac{m_{W}^{2}}{q^{2}} \right)^{n}
=\frac{1}{16\pi^{2}}\left(1-\frac{m_{W}^{2}}{m_{t}^{2}}\right)\pi
\end{equation}
is in agreement with the one in Sec.~2.1.3.


\subsection{Next-to-leading order (NLO) QCD corrections}

The traditional technique used for NLO calculation is to calculate the one-loop
and tree-graph contributions separately. In the present case the UV 
singularities are regularized 
by dimensional regularization whereas the IR/M singularities are regularized
by introducing  gluon and bottom quark masses. The IR/M singularities will
eventually appear as $(\ln m_{g})-$ and $(\ln m_{b})-$ singularities and cancel
among the one-loop and tree graph contributions
\cite{Fischer:2001gp,Do:2002ky,Fischer:1998gsa,Jezabek:1988iv,
Fischer:2000kx}.
We mention that the calculation can also be done in dimensional regularization
without recourse to the traditional $m_{g}\neq 0$ and $m_{b}\neq 0$ 
regularization \cite{Czarnecki:1990kv}.

For example, generic diagrams for the QCD NLO calculation are displayed in 
Fig.~5.
\begin{figure}[htbp]
\begin{center}
\begin{minipage}[t]{5cm}
\includegraphics[width=5cm]{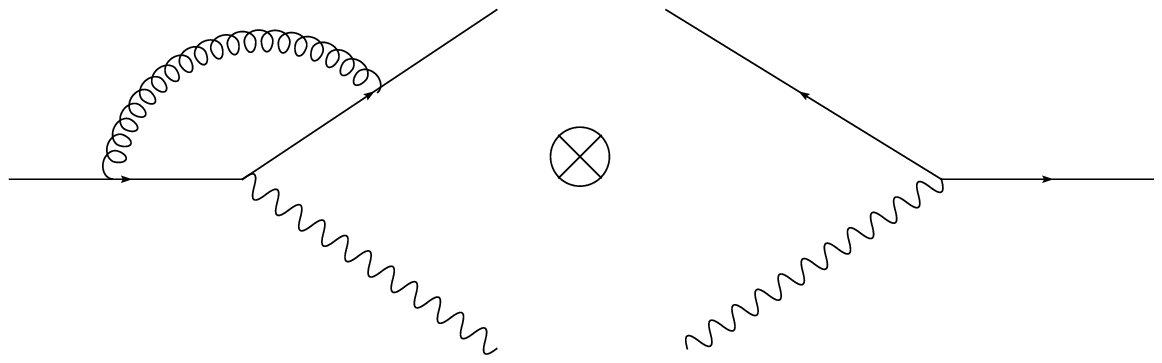}
\end{minipage}
\begin{minipage}[t]{6cm}
\includegraphics[width=6cm]{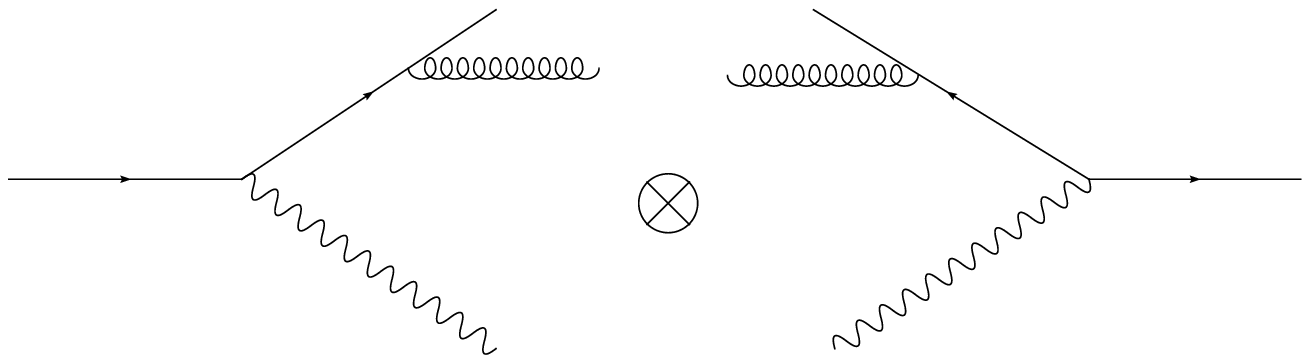}
\end{minipage}
\put(-235,55){\small 1--loop $\otimes$ Born}
\put(-90,55){\small tree $\otimes$ tree}
\caption{Generic NLO QCD contributions}
\end{center}
\end{figure}
Without going into the details of the calculation 
(see e.g.~\cite{Fischer:2001gp}) we just
quote the result of the NLO calculation.
For the total rate one obtains ($\hat{\Gamma}=\Gamma/\Gamma({\rm Born})$)
(see e.g.~\cite{Fischer:2000kx})
\begin{eqnarray}
\label{UplusL}
\hat{\Gamma}(NLO) & = & 1 \!+\!
  \frac{\alpha_s} {2 \pi} C_F \frac{y^2}{(1 \!-\! y^2)^2 (1 \!+\! 2 y^2)}
  \Bigg\{ \frac{(1 \!-\! y^2)(5 \!+\! 9 y^2 \!-\! 6 y^4)}{2 y^2} \!-\!
  \frac{2 (1 \!-\! y^2)^2 (1 \!+\! 2 y^2) \pi^2}{3 y^2}
  \nonumber \\ & & \hspace{-0.96cm} \,-\,
  \frac{(1 \!-\! y^2)^2 (5 \!+\! 4 y^2)}{y^2} \ln (1 \!-\! y^2) \!-\! 
  \frac{4 (1 \!-\! y^2)^2 (1 \!+\! 2 y^2)}{y^2} \ln(y) \ln(1 \!-\! y^2) \!-\!
  4 (1 \!+\! y^2) \!\times\! \hspace{5mm} \nonumber \\ & & \hspace{-0.96cm} \,\times\, 
  (1 \!-\! 2 y^2) \ln(y) - \frac{4 (1 \!-\! y^2)^2 (1 \!+\! 2 y^2)}{y^2}
  \mbox{Li}_2(y^2) \Bigg\}\,. 
\end{eqnarray}

The numerical value of the NLO QCD correction appears in 
Eq.(\ref{ratenum}). Our numerical input values are 
$m_{t}=175$ GeV and $m_{W}=80.419$ GeV. The strong 
coupling constant has been evolved from $ \alpha_s(M_Z) = 0.1175 $ to 
$ \alpha_s(m_t) = 0.1070 $ using two-loop running.
Numerically one has $\Gamma= \Gamma({\rm Born})(1-8.54 \%)$. 
One sees that the NLO QCD corrections reduce the Born 
term rate by the large amount of $8.5 \%$.

In the limit $y \to 0$ one obtains
\begin{equation}
\label{yto0}
\hat{\Gamma}(\text{NLO}) = 1 +
\frac{\alpha_s} {2 \pi} C_F \left\{\frac{5}{2}-\frac{2}{3}\pi^{2}\right\}\,¸,.
\end{equation} 
The leading $y \to 0$ contribution reduces the rate by $9.26\%$ which is 
already quite close to the rate reduction of the full result ($8.5\%$). We 
shall return to an assessment of the
quality of the $y$-expansion later on. It is curious to note that the
radiative QCD corrections reduce the LO rate whereas the radiative QCD 
corrections to the decay
$Z\to q \bar{q}$ enhance the LO rate ratio
$\hat{\Gamma}(\text{LO})$ by 
$\frac{\alpha_s} {2 \pi} (6/4)C_F=\alpha_{s}/\pi $, i.e. 
$\hat{\Gamma}(\text{NLO};Z\to q \bar{q})=1+\alpha_{s}/\pi$.

The NLO rate can also be calculated by the optical theorem method using the
$y$-expansion. At NLO one has contributions both from the soft and the hard
region leading to an infinite power series in $y$ and $y \ln y$ where the
$(y \ln y)$-contributions come from the interplay of the soft and hard
integration regions. The results of the $y$-expansion have been checked
against the exact result Eq.(\ref{UplusL}) up to $O(y^{16})$
~\cite{Piclum:2008zz} (see also \cite{Blokland:2004nd}).

\subsection{NLO electroweak corrections}
In Fig.~6 we have drawn the LO diagram and the four NLO tree-level diagrams 
that contribute to $ t \rightarrow b + W^+ +(\gamma) $. We use the 
Feynman-'tHooft gauge so that
one has a NLO contribution from the charged unphysical Higgs boson $\chi^{+}$
as shown in Fig.~6. Compare the number of four electroweak NLO tree-level 
diagrams with the two QCD NLO tree-level diagrams. When squaring the
tree-level diagrams one would 
expect a four-fold complexity factor when going from QCD to the electroweak
tree-graph corrections. It is therefore quite remarkable that the squared 
tree graph expressions in both cases are similar in length and 
structure~\cite{Do:2002ky}.   
\begin{figure}
\begin{center}  
  \psfig{file=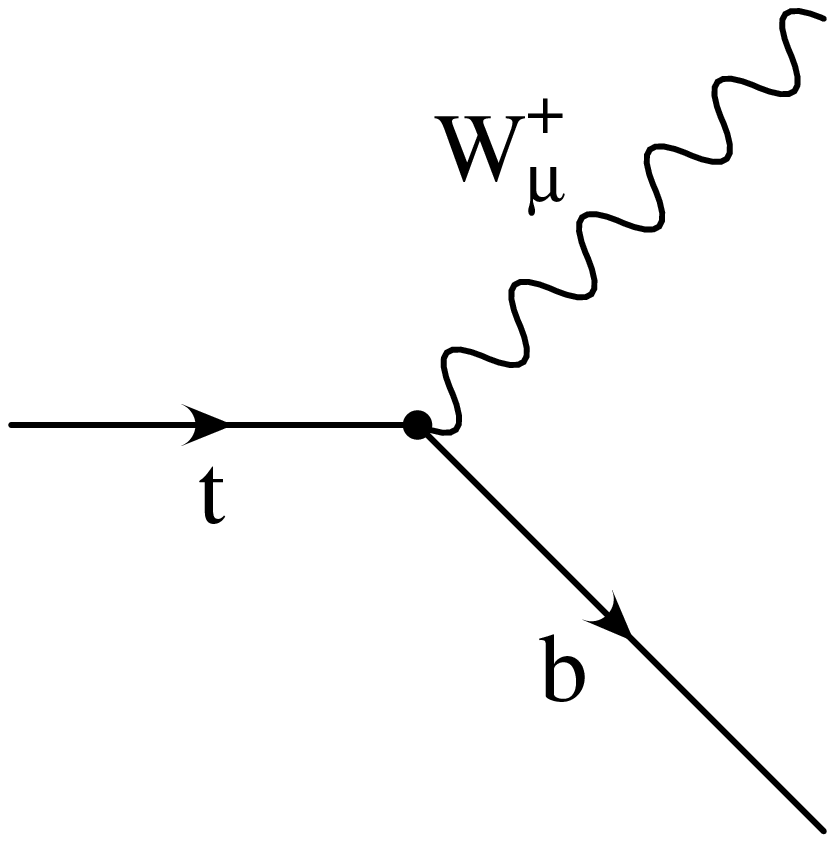,width=2.6cm}
  \psfig{file=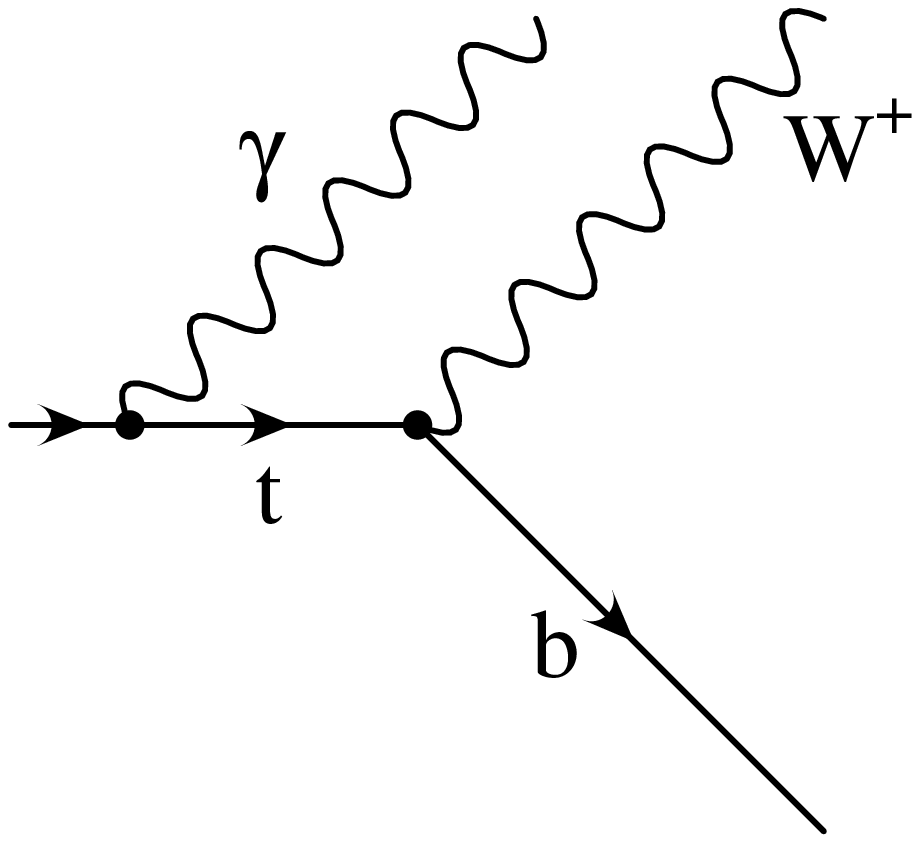,width=2.6cm}
  \psfig{file=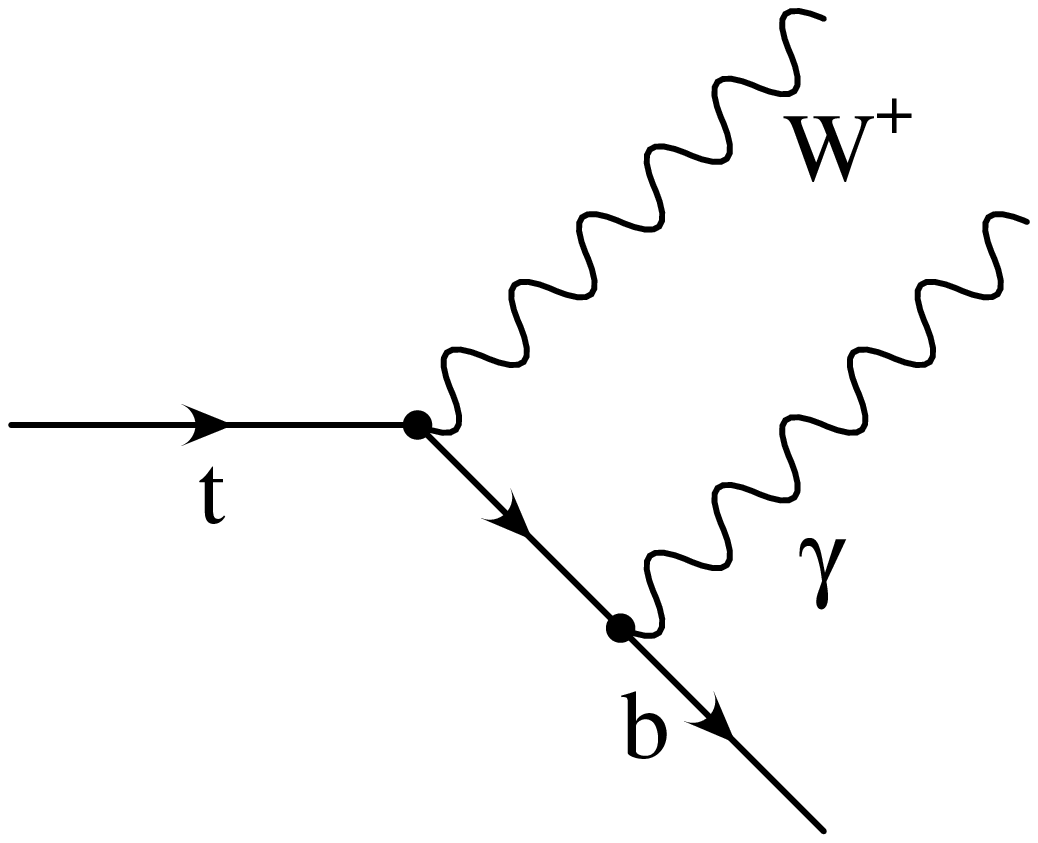,width=2.6cm}
  \psfig{file=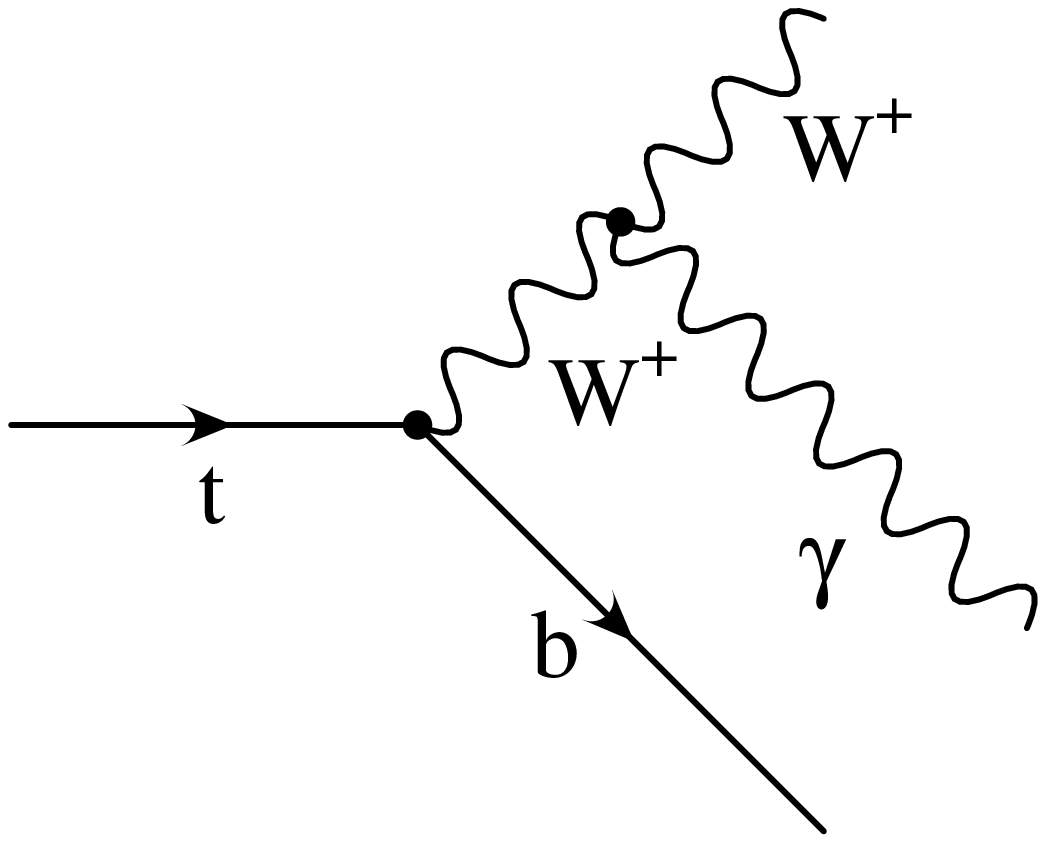,width=2.6cm}
  \psfig{file=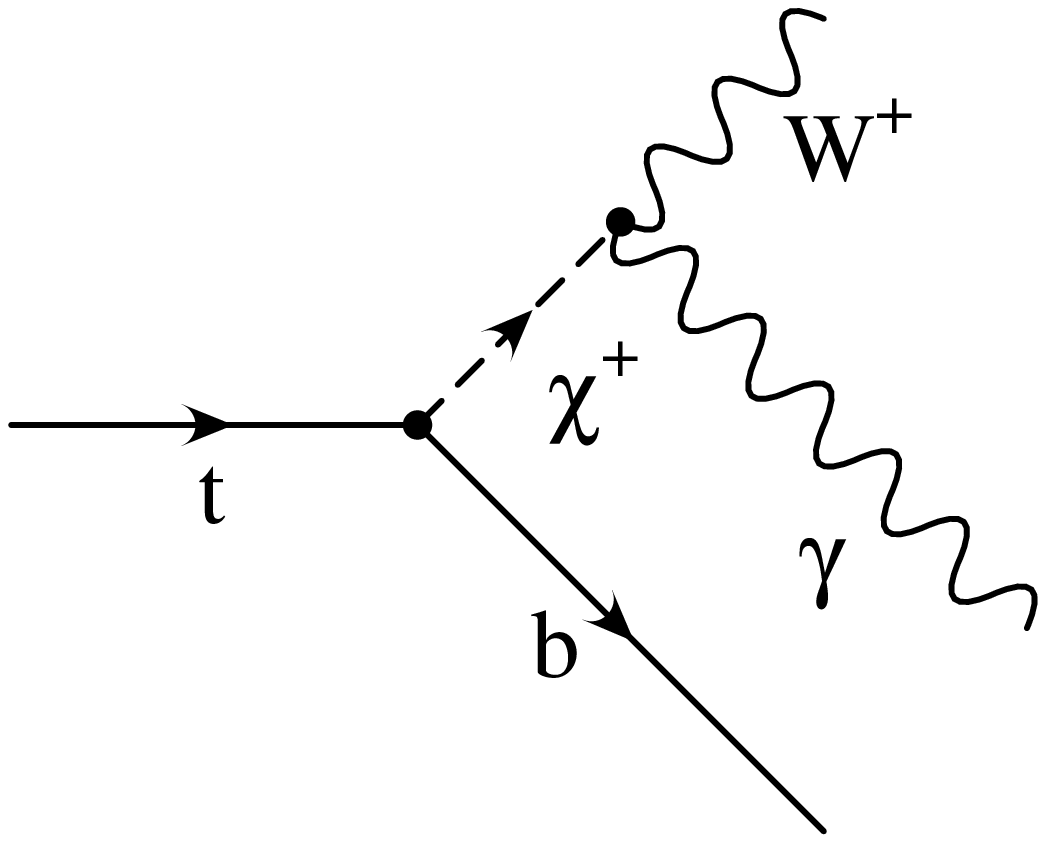,width=2.6cm}
  \caption{Born and electroweak tree-graph contributions to
  $ t \rightarrow b + W^+\, (\gamma) $. $ \chi^+ $ denotes the charged 
Goldstone boson.}
\end{center}
 \end{figure}

\begin{figure}[!htb]
\centering
\includegraphics[width=110mm]{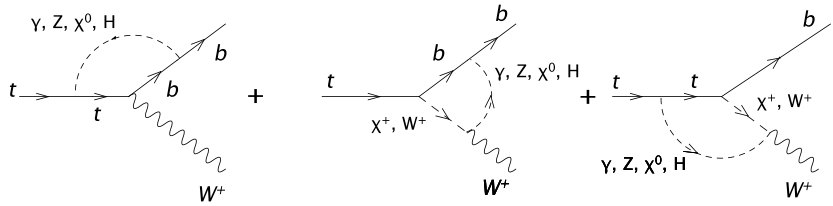}
\caption{Eighteen electroweak three-point one-loop graphs in Feynman-'tHooft 
gauge contributing to $ t \rightarrow b + W^+ $. }
\end{figure} 
In addition to the tree graph contributions one has to consider 18 
three-point one-loop graphs in the Feynman-'tHooft gauge as shown in Fig.~7.
Looking at Fig.~7 one would superficially expect 4+8+8=20 one-loop 
contributions. However, since there is no $(W^{+}W^{+}\chi^{0})$--vertex, this
number reduces to 18 as stated before. In Fig.~7 $\chi^{\pm}$ and 
$\chi^{0}$ are the charged and neutral unphysical Goldstone bosons, and $H$
is the physical Higgs. The results of calculating the one-loop contributions 
exist in amplitude form~\cite{Denner:1990ns}. In the course of calculating
the electroweak radiative corrections to the partial helicity rates the
results of~\cite{Denner:1990ns} were recalculated and confirmed by us. In 
particular we checked
the results of~\cite{Denner:1990ns} numerically with the automated loop 
calculation program XLOOPS/GiNaC developed at the University of 
Mainz~\cite{Frink:1997sg,Brucher:1998ec,Bauer:2000cp}. In addition to the 
one-loop three-point
functions one has a large number of one-loop two-point functions needed
in the one-loop renormalization program. Again these have been reevaluated
using XLOOPS/GiNaC. 

We have used the so-called $ G_F $--renormalization scheme for the electroweak
 corrections where $ G_F $, $ M_W $ and $ M_Z $ are used as input 
 parameters. The $ G_F $--scheme is the appropiate
 renormalization scheme for processes with mass scales that are much larger
 than $ M_W $ as in the present case. The electroweak radiative corrections are
 substantially larger in the so-called $ {\alpha} $--scheme  where $ \alpha $,
 $ G_F $ and $ M_Z $ are used as input parameters. The numerical results
of the electroweak corrections to the rate are given in Eq.(\ref{ratenum}).


\subsection{NNLO QCD corrections}

In the NNLO case squaring of the contributing tree and loop diagrams leads to 
the four generic contributions shown in Fig.~8.
\begin{figure}[htbp]
\begin{center}
\begin{minipage}[t]{5.5cm}
\includegraphics[width=5.2cm]{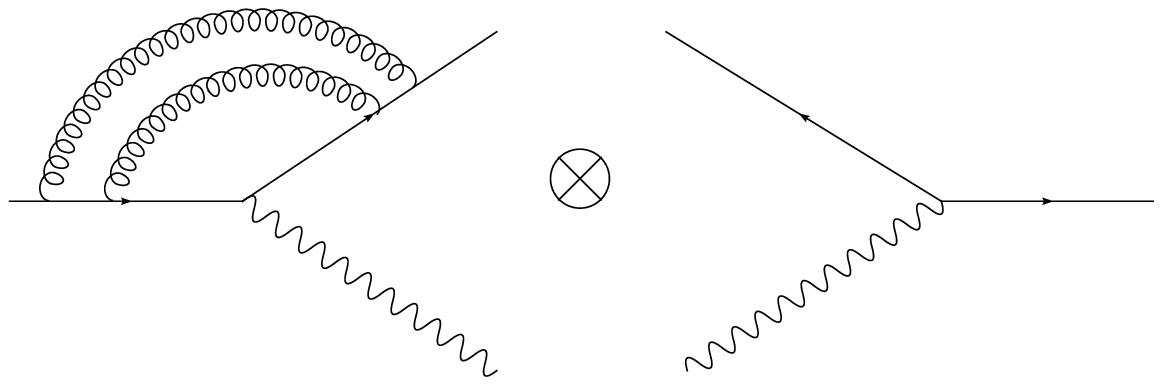}
\end{minipage}
\begin{minipage}[t]{5.5cm}
\includegraphics[width=5.2cm]{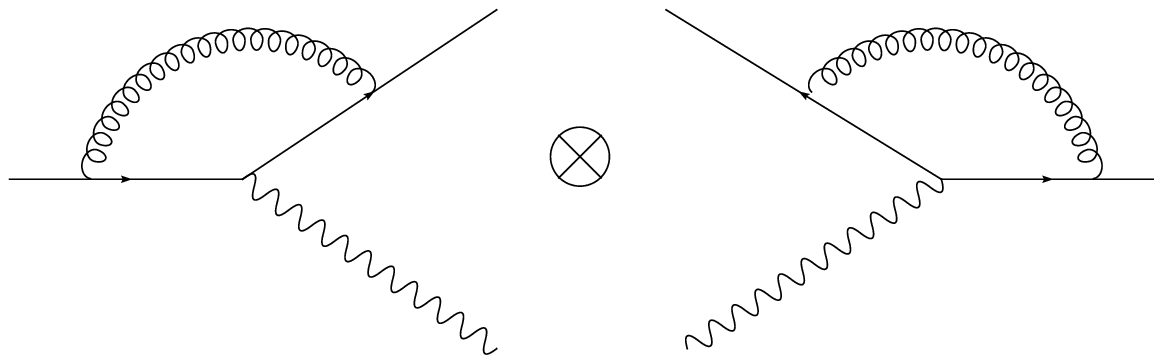}
\end{minipage}
\put(-276,-8){\small 2--loop $\otimes$ Born}
\put(-116,-8){\small 1--loop $\otimes$ 1--loop}
\end{center}
\vspace{0.3cm}
\begin{center}
\begin{minipage}[t]{5.5cm}
\includegraphics[width=5.2cm]{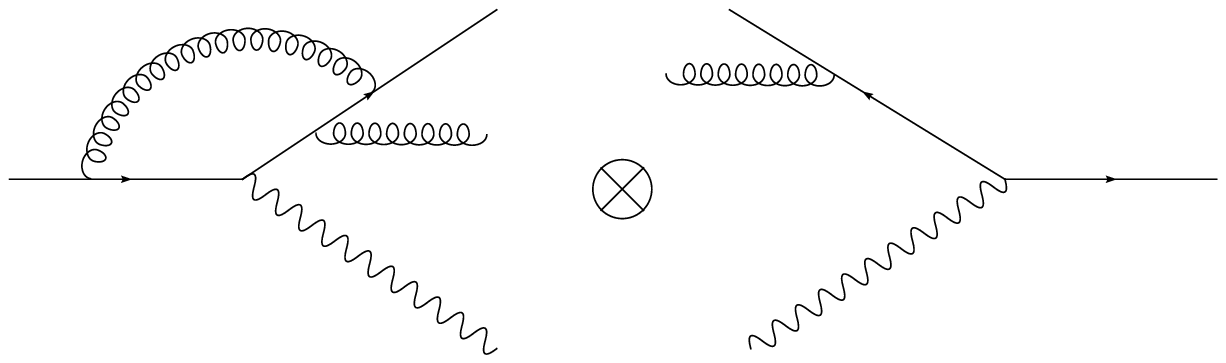}
\end{minipage}
\begin{minipage}[t]{5.5cm}
\includegraphics[width=5.2cm]{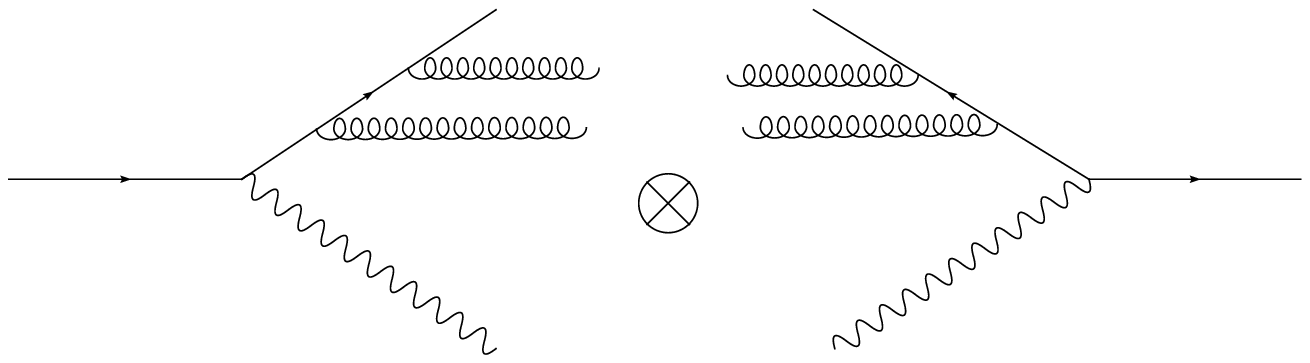}
\end{minipage}
\put(-336,-8){\small 1--loop gluon emission $\otimes$ 1 gluon emission}
\put(-103,-8){\small tree $\otimes$ tree}
\caption{Generic NNLO QCD contributions.}
\end{center}
\end{figure}
However, with present techniques, this method is not viable, mainly because
the NNLO phase space integration become too difficult.

Instead, one resorts again to the optical theorem and calculates the NNLO
rate from the three-loop self-energy diagrams according to 
\cite{Blokland:2004ye,Blokland:2005vq}    
\begin{equation}
  \label{eq::optical}
  \Gamma({\rm NNLO}) =\frac{1}{2s_{t}+1} \frac{1}{m_t}\, {\rm Im} 
 \Sigma(3-loop) \,,
\end{equation}
There are altogether 38 three-loop Feynman diagrams a sample of which are 
shown in Fig.~9.
\begin{figure}[!h]
\begin{center}
  \includegraphics[height=55mm]{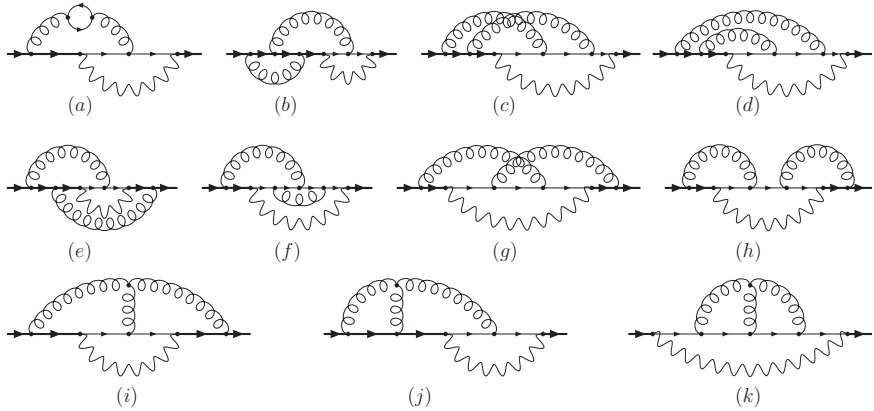}
\caption{Sample three-loop diagrams whose imaginary parts contribute to the 
NNLO calculation of the top quark width.}
\end{center}
\end{figure}

The main ideas of the NNLO calculation of the rate have already been described
in the calculation of the
Born term rate in Sec.~2.1.3. It turns out that again one only has to consider
two momentum regions. In the hard region all loop momenta are hard and the
$W$-propagator can be expanded into a series of massless propagators as in the
LO case. In the soft region the gluon momenta are hard but the loop momentum
flowing through the $W$ is soft. Differing from the LO calculation one now
also has contributions from the soft region. In the soft region the integrals
factorize into 
two-loop self-energy-type integrals and a one-loop vacuum bubble diagram
which are not difficult to integrate. The interplay of the hard and 
the soft region leads to additional $(y^{n}\ln y)$-terms in 
the $y=(m_{W}/m_{t})$-expansion.

One can reduce all integrals to 23 master integrals by 
integration-by-parts identities.
Use was made of Laporta's algorithm in this reduction to master integrals.
The imaginary parts of the master integrals were calculated using the cutting
rules where care had to be taken that some of the master integrals admitted
several ways of cutting them. We mention that 
the calculation had been done in the general covariant gauge
$-g^{\mu\nu}+(\xi-1)k^{\mu}k^{\nu}/k^{2}$\,for the gluon
in order to check on gauge invariance. The numerical results on the
NNLO\,\,QCD corrections are given in Eq.(\ref{ratenum}).


\section{$W$-helicity fractions in top quark decays}


\subsection{Angular decay distribution for 
$t \to b+W^{+}(\to \ell^{+}+\nu_{\ell})\quad(I)$} 
In Fig.~6 we display the LO amplitude contribution to 
$t \to b+\ell^{+}+\nu_{\ell}$. On squaring the amplitude and taking the spin
sums one is led to
the contraction $L_{\mu \nu}H^{\mu \nu}({\rm Born})$.
\begin{figure}[!htb]
\centering
\includegraphics[width=45mm]{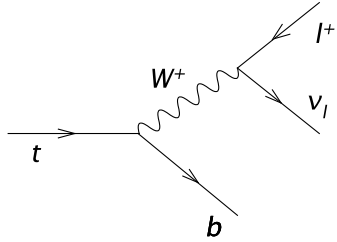}
\caption{LO Born term contribution to $t \to b+W^{+}(\to \ell^{+}+\nu_{\ell})$}
\end{figure}  
For the lepton tensor we obtain
\begin{align}
\label{leptontensor}
L^{\mu \nu}=&\quad\frac{1}{8}{\rm Tr}\,\,\slashed{p}_{\ell}\gamma^{\mu}
(1-\gamma_{5})
\slashed{p}_{\nu}\gamma^{\nu}(1-\gamma_{5}) \nonumber \\
=&\quad p_{\ell}^{\mu}p_{\nu}^{\nu}+p_{\ell}^{\nu}p_{\nu}^{\mu}
-\frac{1}{2}m_{W}^{2}g^{\mu\nu}-i\eps^{\mu \nu\alpha\beta}p_{\ell \alpha}
p_{\nu \beta}\,\,.  
\end{align}
The LO hadron tensor is given by $(m_{b}=0)$
\begin{align}
H^{\mu \nu}({\rm Born})=&\quad\frac{1}{8}{\rm Tr}(\slashed{p}_{t}+m_{t})
\gamma^{\mu}(1-\gamma_{5})
\slashed{p}_{b}\gamma^{\nu}(1-\gamma_{5}) \nonumber \\
=&\quad p_{t}^{\mu}p_{b}^{\nu}+p_{t}^{\nu}p_{b}^{\mu}
-p_{t}\cdot p_{b}g^{\mu\nu}-i\eps^{\mu \nu\alpha'\beta'}p_{t \alpha'}
p_{b \beta'} \,\,.
\end{align}
The factors $1/8$ have been introduced for convenience. The result of
contracting the lepton and hadron tensor reads
\begin{equation}
\label{leptonhadron1}
L_{\mu \nu}H^{\mu \nu}({\rm Born})=4(p_{t}\cdot p_{\ell})\,(p_{b}\cdot 
p_{\nu})=
4(p_{t}\cdot p_{\ell})\,(p_{b}\cdot (q-p_{\ell})\,)\,\,.
\end{equation}
\vspace{0.5cm}
Note that one originally had 
\begin{equation}
\label{leptonhadron2}
L_{\mu\nu}H^{\mu'\nu'}({\rm Born}) (-g_{\mu'}^{\mu}+ \frac{q_{\mu'}q^{\mu}}
{m_{W}^{2}})(-g_{\nu'}^{\nu}+ \frac{q_{\nu'}q^{\nu}}{m_{W}^{2}})
\end{equation}
which turns into $L_{\mu \nu}H^{\mu \nu}({\rm Born})$ in the zero 
lepton mass 
case where
$q_{\mu}L^{\mu\nu}=q_{\nu}L^{\mu\nu}=0$. The lepton mass corrections are of
${\cal O}(m_{\ell}^{2}/m_{t}^{2})$ and are thus negligible. If one wants to 
include lepton mass effects one has to retain the full $W$--projector in 
(\ref{leptonhadron2}). 

One must evaluate the invariant $L_{\mu \nu}H^{\mu \nu}$ in one frame. 
Here we choose the rest frame of the top quark.
Since we want to evaluate $L_{\mu \nu}H^{\mu \nu}$ in terms of the angle
$\cos\theta$ defined in the $W^{+}$-rest frame ($W_{\it r.f.}$) as shown in Fig.~2
we write\footnote{
In Eq.(\ref{pell}) we have specified 
the azimuthal dependence of $p^{\mu}_{\ell}\,(W_{\rm r.f.})$. 
This is not really needed in the present application because we do not 
specify a preferred transverse direction. In general, a transverse direction 
could be defined by the polarization of the top quark or the decay products 
of the $b$-quark. In this case one has to retain the azimuthal dependence of 
the lepton's momentum as done in (\ref{pell}). Whereas the sign of polar 
angle correlations can always be checked by physics arguments, there are
no ready physics arguments to check the signs of the azimuthal correlations. 
To get
the signs of the azimuthal correlations right it is indispensable to use the
boosting method as described above (see e.g.~\cite{Kadeer:2005aq}).} 
\begin{equation}
\label{pell}
p^{\mu}_{\ell}\,(W_{\rm r.f.})=\frac{m_{W}}{2}(1;\sin\theta\cos\phi,
\sin\theta \sin\phi,\cos\theta)\,.
\end{equation}
We then boost the lepton momentum 
$p^{\mu}_{\ell}\,(W_{\rm r.f.})$
to the top quark rest frame ($t_{\rm r.f.}$) where the invariants in 
(\ref{leptonhadron1}) are to be evaluated. The relevant Lorentz boost 
matrix reads
\begin{equation}
\label{boost}
L(boost)=\frac{1}{m_{W}}\left(\begin{array}{cccc}
q_{0} & 0 & 0 & |\vec{q}| \\
0 & m_{W} & 0 & 0 \\
0 & 0 & m_{W} & 0 \\
|\vec{q}| & 0 & 0 & q_{0}
\end{array} \right)
\end{equation}
such that
\begin{equation}
p^{\mu}_{\ell}\,(t_{\rm r.f.})= L(boost)\,\,p^{\mu}_{\ell}\,(W_{\rm r.f.})\,\,.
\nonumber
\end{equation}
The boost will not affect the transverse components $\mu =1,2$ but
only the zero and  longitudinal components $\mu=0,3$. In Eq.(\ref{boost})
$q_{0}$ and $|\vec{q}|$ denote the energy and momentum of the $W$-boson in 
the top quark rest frame.  

In the following we set $m_{b}=0$ such that  
$q_{0}=\frac{m_{t}}{2}(1+y^{2})$ and $|\vec{q}|=\frac{m_{t}}{2}(1-y^{2})$. 
Boosting $p^{\mu}_{\ell}\,(W_{\rm r.f.})$ one obtains
\begin{equation}
p^{\mu}_{\ell}\,(t_ {\rm r.f.})=\frac{m_{t}}{4}
\big((1+y^{2})+(1-y^{2})\cos\theta;\,2y\sin\theta\cos\phi,
2y\sin\theta\sin\phi,(1-y^{2})+(1+y^{2})\cos\theta\big). 
\end{equation}
The remaining  momentum four--vectors in the $t$--rest frame are given by
\begin{eqnarray}
p_{t}^{\mu}&=&m_{t}(1;0,0,0)\,, \nonumber \\
p_{b}^{\mu}&=&\frac{m_{t}}{2}\big(1-y^{2};0,0,-(1-y^{2})\big)\,, \nonumber \\
q^{\mu}    &=&\frac{m_{t}}{2}(1+y^{2},0,0,1-y^{2})\,. 
\end{eqnarray}
We are now in the position to evaluate the invariants 
appearing in Eq.(\ref{leptonhadron1}).
We sort the resulting expression in terms of the
polar angle factors $\sin^{2}\theta$ and $(1\pm\cos\theta)^{2}/2$. Since 
we are not interested in the azimuthal angle dependence in the present 
application we integrate over the azimuthal angle $\phi$. One then obtains the
angular decay distribution
\begin{align}
\label{angdist2}
\int_{0}^{2\pi} d\phi \,\,L_{\mu \nu}H^{\mu \nu}({\rm Born})=2\pi\,
\frac{8}{3}\frac{m_{t}^{4}}{4}
&\bigg\{\quad \underbrace{ \frac{1}{2}(1-y^{2})}_{\sim L}\quad
\frac{3}{4}\sin^{2}\theta \nonumber \\
&\qquad \underbrace{0}_{\sim T_{+}} \quad  \qquad 
\frac{3}{8}(1+\cos\theta)^{2} \nonumber \\ 
&+ \underbrace{y^{2}(1-y^{2})}_{\sim T_{-}} \quad
\frac{3}{8}(1-\cos\theta)^{2} 
\qquad\bigg\}\,, 
\end{align}
where, by comparison with Eq.(\ref{msquared2}), we have identified the three
LO hadron contributions proportional to $L$ and $T_{\pm}$. The normalized
helicity fractions ${\cal G}_{L}$ and ${\cal G}_{\pm}$ written down before in
Eq.(\ref{fractions}) can be read off from Eq.(\ref{angdist2}). As we shall see
later on from an angular momentum analysis, the sorting of the angular 
contributions in (\ref{angdist2}) should be done exactly along the three 
angular factors proportional to $\sin^{2}\theta$ and $(1\pm\cos\theta)^{2}/2$
discussed above.
The corresponding coefficient factors are then proportional to the partial
helicity rates $\Gamma_{L}$ and $\Gamma_{\pm}$,
respectively. An untreated and unsorted  Mathematica output of 
$L_{\mu \nu}H^{\mu \nu}$ would, in general, 
lead to quite lengthy and messy expressions.
  
Repeating the same exercise for $m_{b}\neq 0$ one obtains  
$(x=m_{b}/m_{t})$
\begin{align}
\label{angdist3}
\int_{0}^{2\pi} d\phi \,\,L_{\mu \nu}H^{\mu \nu}({\rm Born})=2\pi\,\frac{8}{3}\frac{m_{t}^{4}}{4}
&\bigg\{
\quad \underbrace{\frac{1}{2}\Big((1-x^{2})^{2}-y^{2}(1+x^{2})
\Big)}_{\sim L} 
\quad\frac{3}{4}\sin^{2}\theta \nonumber \\
&\quad +\underbrace{\frac{1}{2}y^{2}\Big(1-y^{2}+x^{2}
-\sqrt{\lambda}\Big)}_{\sim T_{+}}
\quad \frac{3}{8}(1+\cos\theta)^{2} \nonumber \\ 
&\quad +\underbrace{\frac{1}{2}y^{2}\Big(1-y^{2}+x^{2}+
\sqrt{\lambda}\Big)}_{\sim T_{-}} 
\quad\frac{3}{8}(1-\cos\theta)^{2}   
 \quad \bigg\}\,, 
\end{align}
where
\begin{equation}
\lambda=
\lambda(1,y^{2},x^{2})=1+y^{4}+x^{4}\,-\,2y^{2}x^{2}
-2y^{2}-2x^{2}\,\,.
\end{equation}
For the $m_{b}\neq 0$ normalized helicity fractions one now obtains
\begin{align}
\label{helfrac}
{\cal G}_{L}=& \quad ((1-x^{2})^{2}-y^{2}(1+x^{2}))/N 
\,,\nonumber\\
{\cal G}_{+}=& \quad y^{2}(1-y^{2}+x^{2}-\sqrt{\lambda})/N 
\,,\nonumber \\
{\cal G}_{-}=& \quad y^{2}(1-y^{2}+x^{2}+\sqrt{\lambda})/N\,,  
\end{align}
where
\begin{equation}
N=(1-x^{2})^{2} +y^{2}(1-2y^{2}+x^{2}) \,\,.
\nonumber
\end{equation}
Let us compare the resulting numerical $m_{b}\neq 0$ values for the 
normalized helicity fraction with their $m_{b}=0$ counterparts. One obtains
(we take $m_{b}=4.8\, {\rm GeV}$ as default value)
\begin{alignat}{6}
m_{b}&=4.8 \,{\rm GeV} &:\qquad \qquad &{\cal G}_{L}:{\cal G}_{+}:{\cal G}_{-}
&=\quad 0.7025 &\ :\ & 0.0004 &\ :\ & 0.2971 \,,\nonumber\\
m_{b} &= 0 &: \qquad \qquad &{\cal G}_{L}:{\cal G}_{+}:{\cal G}_{-}&=
\quad 0.7031 &\ :\ & 0 &\ :\ & 0.2969 \,\,.
\end{alignat}
The effect of including the nonvanishing bottom quark mass can be seen to be 
quite small.

Although we have derived the decay distributions (\ref{angdist2}) and
(\ref{angdist3}) for the Born term case, the angular structure is quite general
as will be shown in the next subsection. In the general case one has to 
replace the LO Born
term structure function $H^{\mu \nu}({\rm Born})$ in (\ref{angdist2}) and
(\ref{angdist3}) by their generalized counterparts as e.g. the corresponding 
NLO or NNLO structure functions.
\subsection{Angular decay distribution for 
$t \to b+W^{+}(\to \ell^{+}+\nu_{\ell})$ (II)} 
The $\cos\theta$ dependence of $L_{\mu\nu}H^{\mu\nu}$ can also be worked out 
in a more systematic way by 
using  the completeness relation for the polarization four--vectors
Eq.(\ref{completenessboson}) \footnote{Since the method is general we can 
omit the LO specification in $H^{\mu \nu}({\rm Born})$.}. 
One can then rewrite the contraction of the lepton and hadron 
tensors $L_{\mu\nu}H^{\mu\nu}$ as
\begin{eqnarray}
\label{leptonhadron3}
L_{\mu\nu}H^{\mu\nu} &=& L^{\mu'\nu'}g_{\mu'\mu}g_{\nu'\nu}H^{\mu\nu}
\nonumber\\
&=&\sum_{m,m'} L^{\mu'\nu'}\epsilon_{\mu'}(m)\epsilon_{\mu}^*(m)
\epsilon_{\nu'}^*(m')\epsilon_{\nu}(m')H^{\mu\nu}
\nonumber \\
&=&\sum_{m,m'}
\bigg( L^{\mu'\nu'}\epsilon_{\mu'}(m)\epsilon_{\nu'}^*(m') \bigg) \;
    \bigg(H^{\mu\nu}\epsilon_{\mu}^*(m)\epsilon_{\nu}(m') \bigg) \nonumber \\ 
&=&\sum_{m,m'} L_{mm'} H_{mm'}\,\,.
\end{eqnarray}
We have thereby converted the invariant contraction $L_{\mu\nu}H^{\mu\nu}$
into a contraction over the spatial spherical components $L_{mm'} H_{mm'}
\,(m,m'=+,0,-)$,
where the spatial spherical components of the lepton and hadron tensors
are defined by 
\begin{align}
\label{LandH}
L_{mm'}&=L^{\mu\nu}\epsilon_{\mu}(m)\epsilon_{\nu}^*(m')\,,\nonumber\\
H_{mm'}&=H^{\mu\nu}\epsilon_{\mu}^*(m)\epsilon_{\nu}(m')\,. 
\end{align}
We have again dropped the $q^{\mu}q^{\nu}$-terms in the completeness relation 
in
Eq.~(\ref{completenessboson}) since $q_{\mu}L^{\mu'\nu'}=q_{\nu}L^{\mu'\nu'}=0$
for massless leptons. 
The nice feature of the representation (\ref{leptonhadron3}) is that the left bracket and the 
right bracket in the next
to last row of (\ref{leptonhadron3}) are separately Lorentz invariant. One can
therefore evaluate the left bracket in the $W^{+}$ rest frame, and the right 
bracket in the $t$--rest system without involving any boost.

Let us now specify the the $W^{+}$-rest frame four-vectors that are needed in 
the $W^{+}$-rest frame evaluation of the lepton matrix $L_{mm'}$. 
In the $W^{+}$ rest frame one has 
\begin{eqnarray}
\label{leptonmom1}
p_\ell^\mu &=& {m_{W}}/{2}\, (1; \phantom{-} \sin\theta \cos\phi,
\phantom{-}\,\sin\theta \sin\phi\,,\phantom{-}\cos\theta)\,, \nonumber \\
p_\nu^\mu &=& {m_{W}}/{2}\, (1; -\sin\theta\cos\phi,
\,-\sin\theta \sin\phi\,,\,-\cos\theta)  \,,
\end{eqnarray}
and the polarization vectors \,(in our convention $a^{\mu}=(a_{0},\vec{a})$ 
and $a_{\mu}=(a_{0},-\vec{a})$)
\begin{eqnarray}
\label{polvec2}
\epsilon_{\mu}(L)&=&(0;0,0,-1)\,, \nonumber \\
\epsilon_{\mu}(\pm)&=&\frac{1}{\sqrt{2}}(0;\pm 1,i,0)\,. 
\end{eqnarray}
It is then straight-forward to evaluate 
$L_{mm'}=L^{\mu\nu}\epsilon_{\mu}(m)\epsilon_{\nu}^*(m')$ using the lepton
tensor (\ref{leptontensor}).

The various components of the lepton matrix $L_{mm'}$ can be written in
a very compact and suggestive way in terms of Wigner's small $d^{1}$-function.
One has
\begin{equation}
\label{leptonmatrix}
L_{mm'}(\theta,\phi)=m_{W}^{2}\,\,d^{1}_{m\,1}(\theta)d^{1}_{m'\,1}(\theta)e^{i(m-m')\phi}\,,
\end{equation}
where the spin one $d^{1}$ function is given by (convention of Rose)
\begin{equation}
\hspace{1cm} d^1_{mm'}(\theta)=\left(
\begin{array}{ccc}
\frac{1}{2}(1+\cos\theta) & -\frac{1}{\sqrt{2}}\sin\theta &\frac{1}{2}(1-\cos\theta) \\
\frac{1}{\sqrt{2}}\sin\theta & \cos\theta & -\frac{1}{\sqrt{2}}\sin\theta \\
\frac{1}{2}(1-\cos\theta) & \frac{1}{\sqrt{2}}\sin\theta &\frac{1}{2}(1+\cos\theta) 
\end{array}
\right)\,\,.
\vspace{0.2cm}
\end{equation}
The rows and columns are labeled in the order $(+1,0,-1)$\,. The representation
(\ref{leptonmatrix}) should be of no surprise to anyone who is familiar with
the behaviour of angular momentum states under a rotation by the angles
$\theta$ and $\phi$. In the lepton system $(x',y',z')$ the only nonvanishing
component of the lepton matrix is $L_{+1,+1}=m_{W}^{2}/2$ as the antilepton and
the neutrino are both left-handed (see Fig.~2). Eq.(\ref{leptonmatrix}) 
represents the rotation of the lepton matrix from the lepton system 
$(x',y',z')$ to the hadron system $(x,y,z)$. In the case $m_{l}\neq 0$ one has
to augment Eq.(\ref{leptonmatrix}) by temporal spin $0$ components and
interference contributions of the temporal spin $0$ and spatial spin {1}
components~\cite{Korner:1989qb}.   

When integrating $L_{\mu\nu}H^{\mu\nu}$ over the azimuthal angle $\phi$ one 
remains only with the three diagonal elements of $H_{mm'}$. One has
\begin{align}
\label{LH}
\int d\phi \,\, L_{\mu\nu}H^{\mu\nu}&=
2\pi\, m_{W}^{2}\sum_{m=+1,0,-1}d^{1}_{m,+1}(\theta)
d^{1}_{m,+1}(\theta)H_{mm} \nonumber \\
&=2\pi \frac{2}{3}m_{W}^{2}
\left( \frac{3}{4}\sin^{2}\theta \,
H_{00}+\frac{3}{8}(1+\cos\theta)^{2}H_{++}
+\frac{3}{8}(1-\cos\theta)^{2}H_{--}\right)\,. 
\end{align}
By convention one drops one of the double indices in the diagonal elements
of the hadronic density matrix $H_{mm}$, i.e. one replaces $H_{00}\to H_{0}$ 
and $H_{\pm\pm} \to H_{\pm}$ as has been done in the rest of this paper.
For the LO case one reproduces Eq.(\ref{msquared2}) using
$H_{00}(=H_{L})=|H_{\frac{1}{2};-\frac{1}{2}0}|^{2}
+|H_{-\frac{1}{2};\frac{1}{2}0}|^{2}$, 
\,$H_{++}(=H_{+})=|H_{\frac{1}{2};\frac{1}{2}1}|^{2}$ and
$H_{--}(=H_{-})=|H_{-\frac{1}{2};-\frac{1}{2}-1}|^{2}$ from (\ref{helamp}).

The advantage of method II is that the method can easily be applied to more 
complex decay processes involving spin. Also one can easily incorporate lepton
mass effects and include polarization effects of initial and final state 
particles~\cite{Korner:1989ve,Korner:1989qb}.
For example, method II was applied to the full angular analysis of 
$B\to D,D^{*}+\ell+\nu_{\ell}\,\,(\ell=e,\mu,\tau)$ 
\cite{Korner:1989ve,Korner:1989qb} and the rare decays 
$B \to K,K^{*}+\ell^{+}+\ell^{-}\,\,(\ell=e,\mu,\tau)$ \cite{Faessler:2002ut}
including results on the polarisation of the final lepton. 
Another example is \cite{Kadeer:2005aq} where we have used method II to
describe the semileptonic decay process of a polarized $\Xi^0$,\,\, 
$\Xi^0(\uparrow) \to 
\Sigma^+ + l^- + \bar{\nu}_l$ \,($l^-=e^-, \mu^-$) 
followed by the nonleptonic
decay $\Sigma^+ \to p + \pi^0$. 
In this process the mass difference $M_{\Xi^0}- M_{\Sigma^+}=125.46\,{\rm MeV}$
is comparable to the $\mu$--mass which makes inclusion of
lepton mass effects mandatory. In fact one finds 
$\Gamma(\mu)/\Gamma(e)\approx 1/120$ in this process.
A cascade type
analysis as used in the method II is ideally suited for Monte Carlo 
event generators that describe
complex cascade decays involving particles with spin. In fact, we wrote
a Monte Carlo generator for the above semileptonic $\Xi^0$ decay process 
\cite{Kadeer:2005aq} which was 
profitably used in the analysis of the NA48 data on this process.  


\subsection{Experimental results on helicity fractions}

An early MC study quotes experimental sensitivities of 
$\delta{\cal G}_{L}=0.7\,\%$ and $\delta{\cal G}_{+}=0.3\,\%$ for an integrated
luminosity of 100 ${\rm fb}^{-1}$ at Tevatron II energies which corresponds to 
$\approx 8\cdot 10^{6}$ $(t\bar{t})$-pairs~\cite{Simmons:2000hr}. Compare 
this to the NLO QCD
changes $\delta{\cal G}_{L}=0.7\,\%$ and $\delta{\cal G}_{+}=0.1\,\%$
to be discussed later on which shows that the radiative corrections are of the
same order as the experimental sensitivities.
Much higher
event rates can be reached at the LHC in one year. 
A more recent MC study based on 10 $fb^{-1}$ at the LHC quotes measurement 
uncertainties of $\delta{\cal G}_{L}=1.9\,\%$, $\delta{\cal G}_{+}=0.22\,\%$
and $\delta{\cal G}_{-}=1.8\,\%$~\cite{AguilarSaavedra:2007rs}.

Experimentally, there has been a continuing interest in the measurement
of the helicity fractions. Latest measurements are 
\begin{eqnarray}
&&{\rm CDF} (2008~\cite{Aaltonen:2008ei})\,: \hspace{1.0cm}
{\cal G}_{L}=0.66 \pm 0.16({\rm stat})\pm0.05({\rm syst}) 
\nonumber\\
&&\hspace{3.8cm}{\cal G}_{+}=-0.03 \pm 0.06 ({\rm stat})\pm0.03({\rm syst})
\nonumber   \\
&&{\rm DO} (2009~\cite{d02008})\,:\;\;\;\; 
\hspace{0.9cm}{\cal G}_{L}=0.490 \pm 0.106 ({\rm stat}) 
\pm 0.085 ({\rm syst})  \nonumber \\
&&\hspace{3.8cm}{\cal G}_{+}=-0.104 \pm 0.076 ({\rm stat})\pm0.066({\rm syst})
\end{eqnarray}
All of these measurements are well within the SM 
predictions.


\section{Construction of covariant helicity projectors}


In Eq.(\ref{LandH}) we have defined the helicity structure functions
$H_{m}\, ({m=L,+,-})$ which multiply the angular factors in the
angular decay distribution. According to their definition in 
Eq.(\ref{LandH})
the helicity structure functions $H_{m}$ can be calculated in a frame-dependent
way by use of the frame-dependent polarization vectors (\ref{polvec1}).
It is much more convenient to calculate the helicity structure functions
covariantly, and, in fact, a covariant projection is indispensable for the 
NLO and NNLO calculations. The covariantization is achieved by defining 
covariant helicity projectors
$\IP^{\mu\nu}_m$ which covariantly project onto the helicity structure 
functions via 
\begin{equation}
H_{m}= \IP^{\mu\nu}_m H_{\mu\nu} \hspace{1.0cm}(m=L,+,-). 
\end{equation}
This definition holds for any general hadron tensor structure irrespective 
of the fact that we have dealt only with the Born term hadron tensor up to now.
To construct the covariant helicity projectors we start with their
representation in terms of the $t$-rest frame polarization vectors 
(\ref{polvec1}) according to the definition Eq.(\ref{LandH}).
One has
\begin{eqnarray}
\label{epshelproj}
\IP^{\mu\nu}_L&=&\epsilon^{\ast \mu}(L)\,\epsilon^{ \nu}(L)\,, \nonumber \\
\IP^{\mu\nu}_{\pm}&=&\epsilon^{\ast \mu}(\pm)\,\epsilon^{ \nu}(\pm). 
\end{eqnarray}
In covariantizing the forms $(\ref{epshelproj})$ it helps to remember that
the helicity projectors must be four-transverse to the momentum
of the $W^{+}$, i.e. they must satisfy
\begin{equation}
q_{\mu}\IP^{\mu\nu}_m=q_{\nu}\IP^{\mu\nu}_m=0\,\,.
\end{equation}
Further, they must satisfy the orthonormality and completeness relations
\begin{alignat}{4}
\label{ortho}
&{\rm orthonormality}\,&:&\quad g_{\mu\nu}\IP^{\mu\nu}_{m} &=& -1 \nonumber \\
&                    & &\quad g_{\alpha \beta}\IP^{\mu\alpha}_{m}\IP^{\beta \nu}_{n} 
 &=& -\delta_{mn} \IP^{\mu\nu}_{m}  \nonumber\\
&{\rm completeness}  &:&\quad \sum_{m} \IP_{m}^{\mu\nu} 
 &:= &\, \IP_{U+L}^{\mu\nu}= -g^{\mu\nu}+\frac{q^{\mu}q^{\nu}}{m_{W}^{2}} 
\end{alignat}
As it turns out the covariant projectors can be constructed from the 
following three projectors
\begin{itemize}
\item Projector for the total rate $\IP^{\mu\nu}_{U+L}$
\begin{equation}
\label{pUplusL}
    \IP^{\mu\nu}_{U+L} = - g^{\mu\nu} + \frac{q^{\mu} q^{\nu}}
{m_W^2}
\end{equation}
\item Projector for the longitudinal helicity rate $\IP^{\mu\nu}_L$
\begin{equation}
\label{pL}
 \IP^{\mu\nu}_L = \frac{m_W^2}{m_{t}^{2}}\frac{1}{|\vec q\,|^{2}} 
    \Big( p_t^{\mu} - \frac{p_t \cdot q}{m_W^2} q^{\mu} \Big)
    \Big( p_t^{\nu} - \frac{p_t \cdot q}{m_W^2} q^{\nu} \Big)
\end{equation}
\item Projector for the forward-backward asymmetric helicity rate 
$\IP^{\mu\nu}_F=\IP^{\mu\nu}_{+}-\IP^{\mu\nu}_{-}$\\ $(\epsilon_{0123}=1)$ \\
\begin{equation}
\label{pF}
 \IP^{\mu\nu}_F = \frac{1}{m_t}\frac{1}
{|\vec q\,|}
    i \epsilon^{\mu \nu \alpha \beta} p_{t, \alpha} q_{\beta}
\end{equation}
\end{itemize}
The denominator factor $|\vec q\,|^{2}$ refers to the top quark rest frame.
In invariant form the normalization factor is given by  
$|\vec q\,|^{2}=
((p_{t}q)^{2}-m_{W}^{2}m_{t}^{2})/m_{t}^{2}$\,.
Finally, the three projectors read ($H_{m}= \IP^{\mu\nu}_m H_{\mu\nu};\,\,\,\,m=L,+,-$)

\begin{eqnarray}
\label{covhelproj}
\IP^{\mu\nu}_L &=&\frac{m_W^2}{m_t^2} \frac{1}{|\vec{q}\,|^2} 
    \Big( p_t^{\mu} - \frac{p_t \cdot q}{m_W^2} q^{\mu} \Big)
    \Big( p_t^{\nu} - \frac{p_t \cdot q}{m_W^2} q^{\nu} \Big)\nonumber\,, \\
\IP^{\mu\nu}_{\pm} &=& \frac{1}{2}\left(\IP^{\mu\nu}_{U+L}-\IP^{\mu\nu}_L
\pm \, \IP^{\mu\nu}_F \right) \,\,.
\end{eqnarray}
It is instructive to check that, in the $t$-rest frame or in the $W^{+}$-rest 
frames, the 
covariant helicity projectors in Eq.~(\ref{covhelproj}) reduce to the form 
(\ref{epshelproj}) in terms of the rest frame polarization vectors 
(\ref{polvec1}) and (\ref{polvec2}), respectively. Note, though, that in the
$W$-rest frame the normalization factor $|\vec{q}\,|$ in Eqs.~(\ref{pL}) 
and (\ref{pF}) has to be replaced
by\, $y|\vec{p_{t}}\,|$ where $|\vec{p_{t}}\,|$ is the magnitude of the
top quark momentum in the $W^{+}$-rest frame.

The denominator factors $|\vec q\,|^{-2}$ and $|\vec q\,|^{-1}$ in
Eqs.~(\ref{pL}) and (\ref{pF}) are needed for
the correct normalization of the projectors, {\it cif.} Eq.(\ref{ortho}).
As we shall see later on the denominator factors $|\vec q\,|^{2}$ and 
$|\vec q\,|$ somewhat 
complicate the NLO and NNLO calculation
of the helicity rates as compared to the total rate. 

\section{Narrow width approximation}


Let us begin with by discussing how to factorize
of the three--body rate $\Gamma (t \to b +\ell^{+}+\nu_{\ell})$ into
the two--body rates $\Gamma(t \to b +W^{+})$ and 
$\Gamma(W^{+} \to \ell^{+}+\nu_{\ell})$ using the narrow width approximation
for the $W$-boson. The rate formula for the three body 
decay $t \to b+\ell^{+}+\nu$ reads (see \cite{Peskin:1995ev})
\begin{equation}
\Gamma_{3}=\frac{1}{2m_{t}}
\int \frac{1}{(2\pi)^{3}}\frac{d^{3}p_{b}}{2\,E_{b}}
\int \frac{1}{(2\pi)^{3}}\frac{d^{3}p_{l^{+}}}{2\,E_{l^{+}}}
\int \frac{1}{(2\pi)^{3}}\frac{d^{3}p_{\nu_{l}}}{2\,E_{\nu_{l}}}
\frac{1}{2}|\overline{M_{3}}|^{2}(2\pi)^{4}\delta^{(4)}(p_{t}-p_{b}-p_{\ell^{+}}-p_{\nu_{l}}) \,,
\end{equation}
which we write as 
\begin{equation}
\Gamma_{3}=\frac{1}{2}R_{3}\left[|\overline{M_{3}}|^{2} \right]\,.
\end{equation}
The squared three-body matrix element $|\overline{M_{3}}|^{2}$ is given by
\begin{equation}
\label{melement}
|\overline{M_{3}}|^{2}=64 \frac{g_{\omega}^{2}}{8}L_{\mu\nu}
\frac{g_{\omega}^{2}}{8} |V_{tb}|^{2}H^{\mu\nu}
\Big| \frac{1}{q^2 \!-\! m_W^2 \!+\! i m_W \Gamma_W} \Big|^2\,,
\end{equation}
where we have introduced the Breit-Wigner line shape to account for the
finite width of the $W^{+}$--boson. We have also reinstituted the factor
of $8\cdot8=64$ in (\ref{melement}) which was introduced earlier for 
convenience.

Next we introduce the identity
\begin{equation}
\label{phasespace1}
1=\int dq^{2} \int \frac{d^{3}q}{2\,E_{W}}
\delta^{(4)}(q-p_{\ell^{+}}-p_{\nu_{l}}) \,,
\end{equation}
which can be seen to be true in the $W^{+}$ rest frame where $q^{2}=E_{W}^{2}$
and $\int \frac{dE_{W}^{2}}{2E_{W}}\delta (E_{W}-E_{l^{+}}-E_{\nu_{l}})=~1$.
The identity (\ref{phasespace1}) allows one to
factorize the three-body phase space integral 
$R_{3}(t \to b + \ell^{+}+\nu_{\ell})$ into the two-body phase space integrals
$R_{2}(t \to b + W^{+})$ and $R_{2}(W^{+}\to l^{+}+\nu_{l})$. One has
\begin{eqnarray}
\label{fact1}
R_{3}=2m_{W}\int \frac{dq^{2}}{(2\pi)}&&\overbrace{\frac{1}{2m_{t}} \bigg\{
\int \frac{1}{(2\pi)^{3}}\frac{d^{3}p_{b}}{2\,E_{b}}
\int \frac{1}{(2\pi)^{3}}\frac{d^{3}q}{2\,E_{W}}
(2\pi)^{4}\delta^{(4)}(p_{t}-p_{b}-q) \bigg\}}^{R_{2}(t \to b + W^{+})}
 \nonumber \\
&&\hspace{-1.0cm}\underbrace{\frac{1}{2m_{W}} \bigg\{
\int \frac{1}{(2\pi)^{3}}\frac{d^{3}p_{\ell^{+}}}{2\,E_{\ell^{+}}}
\int \frac{1}{(2\pi)^{3}}\frac{d^{3}p_{\nu_{l}}}{2\,E_{\nu_{l}}}
(2\pi)^{4}\delta^{(4)}(q-p_{\ell^{+}}-p_{\nu_{l}}))\bigg\}}_{R_{2}(W^{+}\to
l^{+}+\nu_{l})}\,\,. 
\end{eqnarray}
The phase space nicely factorizes. But how about the factorization of the
squared three--body matrix element 
$|\overline{M_{3}}|^{2}\,$? 
The matrix element squared also factorizes 
{\it after angular integration} which can be seen by using the
relation
\begin{equation}
\int d\cos\theta \,d\phi L_{mn}(\theta,\phi)=\frac{4\pi}{3}
\,m_{W}^{2}\delta_{mn}
\end{equation}
which follows from the explicit representation of $L_{mn}(\theta,\phi)$
given in
Eq.(\ref{leptonmatrix}).
In fact, one has
\begin{equation}
\label{fact2}
\int d\cos\theta \,d\phi\Big(\sum_{m,n}H_{mn} \sum_{m,n}L_{mn}(\theta,\phi)
\Big)=
\frac{1}{{\rm Tr}\,\{\delta_{mn}\}}\Big(\sum_{n}H_{nn}\Big)
\int d\cos\theta \,d\phi 
\Big( \sum_{m}L_{mm}(\theta,\phi)\Big)\,. 
\end{equation}
The factor $1/{\rm Tr}\,\{\delta_{mn}\}=1/3$ provides for the 
crucial statistical factor $1/(2s_{W}+1)$ in the $W^{+}$ width formula. Note
that the
explicit angular integrations over $\cos \theta$ and $\phi$ appearing in 
(\ref{fact2}) are implicit in (\ref{fact1})
\footnote{We mention that an alternative derivation of 
the appearance of the statistical factor
1/3 has been given in \cite{Uhlemann:2008pm}.}. One thus finds
\begin{eqnarray}
\Gamma(t \to b+\ell^{+}+\nu)&=&\frac{m_{W}}{\pi}\int dq^{2} 
\frac{1}{2}R_{2}\Big[|\overline{M}|^{2}(t \to b + W^{+})\Big] \nonumber \\
&&\cdot\,\frac{1}{3}R_{2}\Big[|\overline{M}|^{2}(W^{+} \to \ell^{+}+\nu)\Big]
\Big| \frac{1}{q^2 \!-\! m_W^2 \!+\! i m_W \Gamma_W} \Big|^2\,\,.
\end{eqnarray}

The narrow--width approximation consists in the replacement of the Breit-Wigner
line shape by a $\delta$--function, ${\it cif.}$
\begin{equation}
\Big| \frac{1}{q^2 \!-\! m_W^2 \!+\! i m_W \Gamma_W} \Big|^2 =
 \frac{\pi}{m_W \, \Gamma_W} \frac{1}{\pi} \frac{m_W \, \Gamma_W}
 {(q^2 \!-\! m_W^2)^2 + m_W^2 \Gamma_W^2} \stackrel{\Gamma_{W} \to 0}
{=}
\frac{\pi}{m_W \, \Gamma_W} \delta(q^2 - m_W^2)\,\,.
\end{equation}
Using the narrow-width approximation for the $W^{+}$-boson the 
three-body decay $t \to b+\ell^{+}+\nu$ can be seen to factorize, ${\it cif.}$
\begin{eqnarray}
\label{fact3}
\Gamma (t \to b +\ell^{+}+\nu_{\ell})&=& \Gamma(t \to b +W^{+})
\frac{\Gamma(W^{+} \to \ell^{+}+\nu_{\ell})}{\Gamma_{W}} \nonumber \\
&=& \Gamma(t \to b +W^{+}) \,\,BR(W^{+} \to \ell^{+}+\nu_{\ell} ) 
\end{eqnarray}
which is a result which one expects from physical intuition. Incidentally, the
derivation
of the factorization formula (\ref{fact3}) was posed as one of the problems
in the 2004 TASI lectures of T.~Han~\cite{Han:2005mu}. Judging from the 
contents of this subsection this was not one of his simpler problems.

The numerical value of the finite-width correction to the total width listed 
in (\ref{ratenum}) 
consists of the replacement of $\delta(q^2 - m_W^2)$ by the
Breit--Wigner line shape and integrating over $q^{2}$, ${\it cif.}$
\begin{equation}
    \int^{m_t^2}_{0} dq^2\, 
   \delta(q^2 - m_W^2) \,\, \rightarrow 
   \int^{m_t^2}_{0} dq^2 \frac{m_W \Gamma_W}{\pi}
   \frac{1}{(q^2-m_W^2)^2 + m_W^2 \Gamma_W^2}
\end{equation}
where $ \Gamma_W $ is the width of the $ W $-boson ($ \Gamma_W = 2.12$ GeV ).

Numerically the finite-width correction to the total Born term rate amounts 
to 1.56\% 
(see Eq.(\ref{ratenum})) and is of 
the order of $\Gamma_{W}/m_{W}=2.64\%$ as would be expected. A more extensive 
discussion on finite width corrections can be found in 
\cite{Uhlemann:2008pm,Calderon:2008zz}.
\section{Higher order corrections to helicity fractions}
\subsection{NLO QCD and electroweak corrections}
As in the calculation of the NLO total rate structure function 
$H_{U+L}=H_{+}+H_{-}+H_{L}$ (we call $H_{+}+H_{-}=H_{U}$ where $U$ stands
for the ``unpolarized transverse'') we have employed 
the traditional technique when
calculating the  helicity structure functions 
$H_{L}$ and $H_{\pm}$, i.e. we have separately calculated the hadronic
loop and tree contributions after contracting them with the relevant
projectors $\IP^{\mu\nu}_L,\,\IP^{\mu\nu}_{\pm}$.

As mentioned before, the appearance of the normalization factors 
$|\vec{q}|^{-1}$ and $|\vec{q}|^{-2}$ in the projectors make the calculation
technically more difficult than that for the total rate. For the one-loop
contribution the additional normalization factors are of no concern since they
appear only as overall factors outside of the one-loop integral. This is
different for the phase space integration of the tree-graph contributions
where the normalization factors appear under the integral.  Typically one of
the phase space integrations is over the scaled invariant mass of the 
bottom quark and the gluon  $z=(p_{b}+p_{g})^{2}/m_{t}^{2}$. The
normalization factors then appear as overall factors $|\vec{q}|^{-1}$ and 
$|\vec{q}|^{-2}$ in the phase space integral, where  
\begin{equation}
|\vec{q}|= \frac{m_{t}}{2}\sqrt{\lambda(1,y^{2},z)}\,\,.
\end{equation}
The ensuing
class of phase space integrals is more general and more difficult than the
class of integrals appearing in the total rate calculation. Nevertheless,
the phase space integrations can still be done in closed form.

As a sample result we present the $m_{b}=0$ result for 
$\hat{\Gamma}_L = \Gamma_{L}/\Gamma_{U+L}({\rm Born})$. One 
obtains~\cite{Fischer:2001gp,Fischer:1998gsa,Fischer:2000kx}
\begin{eqnarray}
\label{nlolong}
\hspace{-1.0cm}
\hat{\Gamma}_{L}({\rm NLO}) & = &\frac{1}{(1 \!-\! y^2)^2 (1 \!+\! 2 y^2)}\Big((1-y^{2})^{2} +
  \frac{\alpha_s} {2 \pi} C_F 
  \Big\{  (1 \!-\! y^2 )(5 \!+\! 47 y^2 \!-\! 4 y^4)/2 
  \nonumber \\ 
 &-& \frac{2 \pi^2}{3} (1 \!+\! 5 y^2 \!+\! 2 y^4) \!-\! 3 (1 \!-\! y^2)^2
  \ln (1 \!-\! y^2) + 16y^{2} (1 \!+\! 2 y^2) \ln(y) -
  2 (1 \!-\! y)^2  \nonumber \\ & \times &
  (2 \!-\! y \!+\! 6 y^2 \!+\! y^3) \ln (1 \!-\! y) \ln(y) - 
  2 (1 \!+\! y)^2 (2 \!+\! y \!+\! 6 y^2 \!-\! y^3) \ln(y)
  \ln (1 \!+\! y)  \nonumber \\ & - &
  2 (1 \!-\! y)^2 (4 \!+\! 3 y \!+\! 8 y^2 \!+\! y^3)
  \mbox{Li}_2(y) - 2(1 \!+\! y)^2 (4 \!-\! 3 y \!+\! 8 y^2 \!-\! y^3)
  \mbox{Li}_2(-y) \Big\}\Big) \,. 
\end{eqnarray}

In the limit $y \to 0$ one finds $\hat{\Gamma}_{L}({\rm NLO})=
1+\alpha_{s}C_{F}/(2\pi)(5/2-2\pi^{2}/3)$ and thus 
$\hat{\Gamma}_{L}({\rm NLO})$ saturates the total rate
$\hat{\Gamma}({\rm NLO})$ (see Eq.(\ref{yto0})) in this limit. This is  
expected since
$\Gamma_{U}/\Gamma_{L} \propto m_{W}^{2}/m_{t}^{2}=y^{2}$. Results for the 
other two
NLO QCD helicity rates $\hat{\Gamma}_{+}$ and $\hat{\Gamma}_{-}$ can be found 
in \cite{Fischer:2001gp,Fischer:1998gsa,Fischer:2000kx}. The NLO electroweak
corrections to the helicity rates can be found in~\cite{Do:2002ky}.

Let us summarize our numerical NLO results on the helicity fractions including
also the finite width corrections discussed in Sec.5. We write
 \begin{equation}
   \Gamma_i =  \Gamma_i({\rm Born})
 + \Delta \Gamma_i({\rm QCD})
 + \Delta \Gamma_i({\rm EW})  +
   \Delta \Gamma_i({\rm FW})
 + \Delta \Gamma_i(m_b\neq 0)\,\,.
 \end{equation}
 As before we normalize the partial rates to the total Born term rate
 $ \Gamma_{U+L}({\rm Born}) $. Thus we
 write $ \hat{\Gamma}_i = \Gamma_i/\Gamma_{U+L}({\rm Born})$ $ (i = +,-,L) $.
 For the transverse-minus and longitudinal rates we factor out the normalized
 partial Born rates $ \hat{\Gamma}_i $ and write ($ i = -,L $)
 \begin{equation}
   \hat{\Gamma}_i = 
   \hat{\Gamma}_i({\rm Born})\Big(1 + \delta_i({\rm QCD})
 + \delta_i({\rm EW}) + 
   \delta_i({\rm FW}) + \delta\Gamma_i(m_b\neq 0)\Big)\,,
\end{equation}
 where $ \delta_i = \Gamma_{U+L}({\rm Born}) \, \Delta \Gamma_i/\Gamma_i({\rm Born}) $.
 Writing the result in this way helps to quickly assess the percentage
 changes brought about by the various corrections.  

 Numerically one has
 \begin{eqnarray}
   \label{rateminus}
   \hat{\Gamma}_{-} & = &
   0.297 \Big(1 - 0.0656 ({\rm QCD}) + 0.0206({\rm EW}) -
   0.0197 ({\rm FW}) - 0.00172(m_b\neq 0)\Big) \nonumber\\ 
   & = &    0.297 (1 - 0.0664)\,,
 \end{eqnarray}
 and
 \begin{eqnarray}
   \label{ratel}
   \hat{\Gamma}_{L} & = &
   0.703 \Big(1 - 0.0951 ({\rm QCD}) + 0.0132 ({\rm EW})-
   0.0138 ({\rm FW}) - 0.00357 (m_b\neq 0)\Big) \nonumber \\ 
   & = & 0.703 (1 - 0.0993)\,\,.
\end{eqnarray}
  
 \noindent It is quite remarkable that the electroweak corrections almost
 cancel the finite width corrections in both cases.

 In the case of the transverse-plus rate the partial Born term rate cannot be
 factored out because of the fact that $ \Gamma_{+}({\rm Born}) $ is zero. In 
this case we present our numerical result in the form
   
 \begin{equation}
   \hat{\Gamma}_{+} = \Delta \hat{\Gamma}_{+} ({\rm QCD})
 + \Delta \hat{\Gamma}_{+}({\rm EW})
 + \Delta \hat{\Gamma}_{+}(m_b \neq 0).
\end{equation}
One has   
 \begin{eqnarray}
   \label{rateplus}
   \hat{\Gamma}_{+} &=& 
   0.000927 ({\rm QCD}) + 0.0000745({\rm EW})+
   0.000358 (m_b\neq 0) \nonumber\\ 
   &=& 0.00136.
 \end{eqnarray}
Note that the finite width correction to the transverse-plus helicity rate is 
zero. Numerically the NLO corrections 
to $ \hat{\Gamma}_{+} $ occur only at the pro mille
level. It is save to say that, if top quark decays reveal a violation of the 
SM left-chiral $ (V-A) $ current structure that exceeds the $ 1\% $ level, the
violations must have a non-SM origin such as e.g. an admixture of a
right-chiral $ (V+A) $ current structure in the decay vertex 
$t \to b + W^{+}$. 


\subsection{Quality of the $(m_{W}/m_{t})$--expansion}

In order to check on the quality of the $y=(m_{W}/m_{t})$-expansion we take
the known closed form NLO result (\ref{nlolong}) for $\hat{\Gamma}_{L}$ and 
expand it in powers of $y^{2}$ and $y^{2} \ln y$. The expansion of the curly 
bracket in (\ref{nlolong}) reads 
\begin{align}
\Gamma_{L}(\alpha_{s})= \Gamma_{0}\,\frac{\alpha_{s}}{2\pi}C_{F}
\Big\{...\Big\}&=\Gamma_{0}\,\frac{\alpha_{s}}{2\pi}C_{F}\Bigg\{
\left(\frac{5}{2}-\frac{2 \pi^2}{3}\right)+
\left(40-\frac{10\pi^{2}}{3}\right)y^{2} \nonumber \\ 
&+\frac{1}{9}\left(119-12\pi^{2}-6\ln y\right)y^{4}  
+\left(-\frac{253}{90}+\frac{10\ln y}{3}\right)y^{6}+ ...\Bigg\}\,\,.
\end{align} 
Note that $\Gamma_{L}(\alpha_{s})\to 0$ as the phase space closes at
$y=1$\,\, ($\mbox{Li}_2(-1)=-\pi^{2}/12$).
In Fig.~7 we show a plot of the $y$-dependence of $\Gamma_{L}(\alpha_{s})$
(in units of $[\Gamma_{0}\,\frac{\alpha_{s}}{2\pi}C_{F}]$) for different
orders of $y^{n}$ and for the full result. All curves start at 
$(5/2-2\pi^{2}/3)=0.459$ for $y=0$. The full result goes to zero at $y=1$
remembering that $\mbox{Li}_2(-1)=-\pi^{2}/12$.
As Fig.~7 shows the quality of the expansion is already quite good at 
$O(y^{6})$ even for large \,$y$-values.

This raises the hope that such a $(m_{W}/m_{t})$-expansion can also be 
usefully employed 
in other contexts. One could think of possible applications of the
NNLO calculation of $t \to b + W^{+}$ discussed earlier (which only exists 
in expanded form) to processes such as 
\begin{itemize}
\item $b \to u+\ell^{-}+\bar{\nu}_{\ell}$
\item $\mu^{-} \to e^{-} +\bar{\nu}_{e}+\nu_{\mu}$ 
\end{itemize}
\noindent extending $q^{2}$ over the whole kinematical range 
$0 \leqslant q^{2} \leqslant (m_{1}-m_{2})^{2}$ in these processes. 

\begin{figure}[!htb]
\begin{center}
\includegraphics[width=85mm]{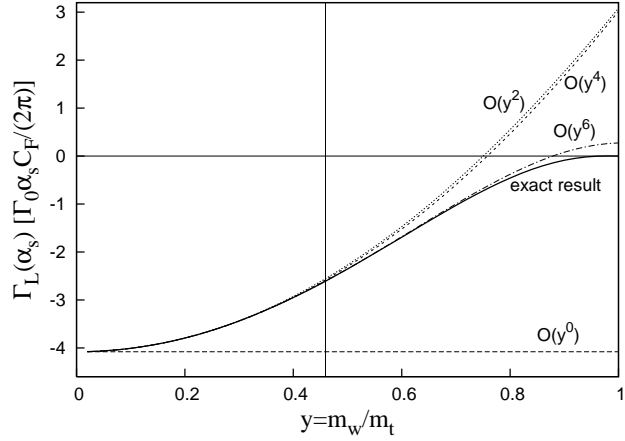}
\end{center}
\caption{Quality of the $y$-expansion of the $\alpha_{s}$ corrections to  
$\Gamma_{L}(\alpha_{s})$. Shown are different orders of the $y$-expansion
in units of $[\Gamma_{0}\,\alpha_{s}\,C_{F}/2\pi]$.
Dashed line: $O(y^{0})$;\, dotted line: $O(y^{2})$;\, dashed line : $O(y^{4})$;
dash-dotted line: $O(y^{6})$;\,\,full line: exact result.   
Vertical line corresponds to the physical point 
$y=m_{W}/m_{t}=0.459\,$. }
\end{figure}

The region very close to the upper kinematical limit of $q^{2}$ given by 
$q^{2}_{max}=(m_{t}-m_{b})^{2}$ requires a separate discussion because this 
region is sensitive to $m_{b}\neq 0$ effects. The upper kinematical 
limit is called the zero recoil point since $\vec{q}=0$ at this
point. For example, at zero recoil $(y=1-m_{b}/m_{t}=1-4.8/175=0.973)$ 
one finds 
\begin{equation}
\label{zrecoil1}
m_{b}\neq 0: \qquad\qquad{\cal G}_{L}:{\cal G}_{+}:{\cal G}_{-}=1/3:1/3:1/3 
\end{equation}
using Eq.(\ref{helfrac}). The equipartitioned helicity fractions result from the fact
that, close to zero recoil, the only surviving transition is the 
allowed Gamow-Teller $s$--wave transition. 
However, for $m_{b}=0$ one has the zero recoil ratios at $y=1$ 
(see Eq.(\ref{fractions}))
\begin{equation}
\label{zrecoil2} 
m_{b}=0: \qquad\qquad{\cal G}_{L}:{\cal G}_{+}:{\cal G}_{-}=
1/3:\phantom{0}0\phantom{0}:2/3\,\,.
\end{equation}

In order to investigate the behaviour of the helicity fractions close to
zero recoil, in Fig.~8 we plot the $y^{2}$-dependence of the helicity 
fractions for $m_{b} \neq 0$ and $m_{b} =0$ with zero recoil values at
$y=1-m_{b}/m_{t}$ and $y=1$, respectively. In the region close to their 
respective
zero recoil points the curves considerably differ from each other. Away from
zero recoil the $m_{b}=0$ and $m_{b}\neq 0$ curves very quickly approach each 
other. Fig.~8 shows that it is safe to use the $m_{b}=0$ approximation for 
$y$-values below $y\approx0.9$. 
\begin{figure}[!htb]
\begin{center}
\includegraphics[width=60mm]{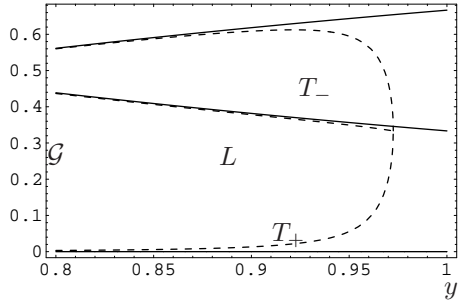}
\put(-90,44){$L$}
\put(-70,15){$T_{+}$}
\put(-60,70){$T_{-}$}
\put(-5,-5){$y$}
\put(-155,45){$\cal G$}
\end{center}
\caption{Helicity fractions ${\cal G}_{L}, {\cal G}_{+}$ and ${\cal G}_{-}$ 
close to 
zero recoil. Dashed line: $m_{b} \neq 0$; full line: $m_{b}= 0$.
The kinematical zero recoil point is given by  
$y=1$ (full line) and $y=1-m_{b}/m_{t}$ (dashed line).}
\end{figure}


\subsection{NNLO QCD corrections to helicity fractions}

In Sec.~2.4 we have desribed how the total NNLO rate can be calculated in a
$y=(m_{W}/m_{t})$-expansion using the optical theorem. Two new features appear 
in the corresponding NNLO calculation of the helicity rates $\Gamma_{L,\pm}$.
First there is a parity violating three-loop contribution which is projected 
out by the projector $\IP^{\mu\nu}_F$. One has to deal with the problem
of how to treat $\gamma_{5}$ in the environment of dimensionally 
regularized loop integrals. We take the prescription 
of~\cite{Larin:1993tq} and replace 
\begin{equation}
\gamma_{\mu}\gamma_{5} \to \frac{1}{3!}\epsilon_{\mu \alpha \beta \gamma}
\gamma^{\alpha}\gamma^{\beta}\gamma^{\gamma}\,\,. 
\end{equation}
When using this prescription one needs to add finite three-loop counter terms 
which are given in~\cite{Larin:1991tj}.

The second new feature is related to the normalization factors
$|\vec q\,|^{-1}$ and $|\vec q\,|^{-2}$ in the three helicity projectors
$\IP^{\mu\nu}_{L,\pm}$ which replace the total rate projector
$\IP^{\mu\nu}_{U+L} = - g^{\mu\nu} + q^{\mu} q^{\nu}/m_W^2$\,.
In the hard region one can expand in inverse powers of the (large) propagator
pole factor $P=(p_{t}+q)^{2}-m_{t}^{2}$. 
\begin{equation}
|\vec q\,|^{2}= q_{0}^{2}-m_{W}^{2}  
= \left(\frac{p_{t}q}{m_{t}}\right)^{2} - m_{W}^{2}\,\,. 
\end{equation}
One expands in the propagator pole factor $P=(p_{t}+q)^{2}-m_{t}^{2}=2p_{t}q+q^{2}$,
i.e. $p_{t}q= \frac{1}{2}P(1-m_{W}^{2}P^{-1})$ where one can replace
$q^{2}$ by $m_{W}^{2}$ since one is cutting through the $W$-line anyhow. One 
then has
\begin{align}
 \frac{1}{|\vec{q}\,|\,^{2}} &= \frac{4 m_t^2}{P^{2}}
  \sum_{n=0}^\infty \left( \frac{2 m_W^2 P^{2}
  - m_W^4 + 4 m_t^2 m_W^2}{P^{2}} \right)^n \,, \nonumber\\
\frac{1}{|\vec{q}\,|} &= \frac{2 m_t}{P}
  \sum_{n=0}^\infty \binom{2n}{n} \left( \frac{2 m_W^2 N 
  - m_W^4 + 4 m_t^2 m_W^2}{4\, P^2}
  \right)^n \,. 
\end{align}
Thus, the additional propagator-like structures from the projectors are
transformed into a scalar on-shell propagator with momentum $p+q$ and
mass $m_{t}$ raised to arbitrary, integer powers. 
This will eventually lead to twelve additional three-loop master integrals 
next to the master integrals appearing in the total rate calculation 
of~\cite{Blokland:2004ye,Blokland:2005vq} whose imaginary parts can again be  
calculated in closed analytical form using the cutting rules. 

In the soft region one cannot perform an expansion of $|\vec{q}\,|$, since
$|\vec{q}\,|^2 = q_0^2 - m_W^2$ and $q_0$ is of order $m_W$ in the soft
region. However, in this region the W boson loop factorizes. Therefore,
one only has to replace the usual one-loop vacuum bubble integrals with
integrals of the type
\begin{equation}
  \int \frac{{\rm d}^dq}{(q^2 - m_W^2)\, (q_0^2 - m_W^2)^n} \,\,, 
\end{equation}
with $n=1/2$ and 1. These integrals are not difficult to evaluate.  

The validity of the treatment of these two new features has been tested 
against the known NLO results up to ${\cal O}((m_{W}/m_{t})^{16})$
~\cite{Piclum:2008zz}. First
results of the NNLO calculation have been published in \cite{Piclum:2008zz}. 
Complete results on the NNLO calculation of the helicity rates will be 
published soon~\cite{pck09}.


\section{Summary and conclusions}

We have discussed some of the properties of the top quark with an emphasis
on the SM decay properties of the top quark. We have defined partial helicity
rates into polarized $W^{+}$-bosons and have derived the resulting angular
decay distribution of $W^{+}_{pol} \to l^{+}\nu$. We have described the LO
calculation of the partial helicity rates using several methods including
also the optical theorem and a $m_{W}/m_{t}$-expansion as a preparation for 
the description of the NNLO calculation of the total rate and the partial 
helicity rates.   
We have summarily described the main features of NLO QCD and electroweak 
corrections to the total width and the partial helicity rates.

We are looking forward to the LHC era with its expected wealth of data
on the top quark and its decay properties. 


\section*{Acknowledgements}

We are grateful to M.A.~Ivanov and H.G.~Sander for helpful discussions. JGK 
would like to thank A.~Czarnecki and J.~Piclum for their collaboration
on the calculation of the NNLO helicity rates.
\begin{footnotesize}




\begin{thebibliography}{99}
\bibitem{Korner:2003zq}
  J.~G.~K\"orner and M.~C.~Mauser,
  ``One-loop corrections to polarization observables,''
  Lect.\ Notes Phys.\  {\bf 647} (2004) 212
  [arXiv:hep-ph/0306082].
\bibitem{Kuhn:1996ug}
  J.~H.~K\"uhn,
  ``Theory of top quark production and decay,''
  arXiv:hep-ph/9707321.
\bibitem{Chakraborty:2003iw}
  D.~Chakraborty, J.~Konigsberg and D.~L.~Rainwater,
  ``Review of top quark physics,''
  Ann.\ Rev.\ Nucl.\ Part.\ Sci.\  {\bf 53} (2003) 301
\bibitem{Bernreuther:2008ju}
  W.~Bernreuther,
  ``Top quark physics at the LHC,''
  J.\ Phys.\ G {\bf 35} (2008) 083001
  [arXiv:0805.1333 [hep-ph]].
\bibitem{Wagner:2008zz}
  W.~Wagner,
  ``Top-quark physics at the Tevatron,''
  Nucl.\ Phys.\ Proc.\ Suppl.\  {\bf 183} (2008) 67.
\bibitem{Incandela:2009pf}
  J.~R.~Incandela, A.~Quadt, W.~Wagner and D.~Wicke,
  ``Status and Prospects of Top-Quark Physics,''
  arXiv:0904.2499 [hep-ex].
\bibitem{Group:2008nq} Tevatron Electroweak Working Group for the CDF 
Collaboration and D0 Collaborations, arXiv:0903.2503 [hep-ex].
\bibitem{Fischer:2001gp}
  M.~Fischer, S.~Groote, J.~G.~K\"orner and M.~C.~Mauser,
  Phys.\ Rev.\  D {\bf 65} (2002) 054036
\bibitem{Do:2002ky}
  H.~S.~Do, S.~Groote, J.~G.~K\"orner and M.~C.~Mauser,
  Phys.\ Rev.\  D {\bf 67} (2003) 091501

\bibitem{Hill:2005zy}
  C.~S.~Hill, J.~R.~Incandela and J.~M.~Lamb,
  Phys.\ Rev.\  D {\bf 71} (2005) 054029
\bibitem{Grossman:2008qh}
  Y.~Grossman and I.~Nachshon,
  JHEP {\bf 0807} (2008) 016
\bibitem{close92} 
F.E.~Close, J.G.~K\"orner, R.J.N.~Phillips, D.J.~Summers,
  J.\ Phys.\ G {\bf 18} (1992) 1716.
\bibitem{Falk:1993rf}
  A.~F.~Falk and M.~E.~Peskin,
  Phys.\ Rev.\  D {\bf 49} (1994) 3320
\bibitem{Fischer:1998gsa}
  M.~Fischer, S.~Groote, J.~G.~K\"orner, M.~C.~Mauser and B.~Lampe,
  Phys.\ Lett.\  B {\bf 451} (1999) 406
\bibitem{Groote:2006kq}
  S.~Groote, W.~S.~Huo, A.~Kadeer and J.~G.~K\"orner,
  Phys.\ Rev.\  D {\bf 76} (2007) 014012
\bibitem{Jezabek:1988iv}
  M.~Jezabek and J.~H.~K\"uhn,
  Nucl.\ Phys.\  B {\bf 314} (1989) 1.

\bibitem{Denner:1990ns}
  A.~Denner and T.~Sack,
  Nucl.\ Phys.\  B {\bf 358} (1991) 46.

\bibitem{Eilam:1991iz}
  G.~Eilam, R.~R.~Mendel, R.~Migneron and A.~Soni,
  Phys.\ Rev.\ Lett.\  {\bf 66} (1991) 3105.

\bibitem{Blokland:2004ye}
  I.~R.~Blokland, A.~Czarnecki, M.~Slusarczyk and F.~Tkachov,
  Phys.\ Rev.\ Lett.\  {\bf 93} (2004) 062001
\bibitem{Blokland:2005vq}
  I.~R.~Blokland, A.~Czarnecki, M.~Slusarczyk and F.~Tkachov,
  Phys.\ Rev.\  D {\bf 71} (2005) 054004
\bibitem{cdf06}
CDF-coll, Conf. Note 8104, available from 
http://www-cdf.fnal.gov/physics/new/top/top.html
\bibitem{cdf08}
T.~Aaltonen {\it et al.}  [CDF Collaboration],
  Phys.\ Rev.\ Lett.\  {\bf 102} (2009) 042001
\bibitem{Carlson:1995ck}
  D.~O.~Carlson and C.~P.~Yuan,
  arXiv:hep-ph/9509208.
\bibitem{Martinez:2002st}
  M.~Martinez and R.~Miquel,
  Eur.\ Phys.\ J.\  C {\bf 27} (2003) 49
\bibitem{Groote:1995yc}
  S.~Groote, J.~G.~K\"orner and M.~M.~Tung,
  Z.\ Phys.\  C {\bf 70} (1996) 281
\bibitem{Groote:2009zk}
  S.~Groote, J.~G.~K\"orner and J.~A.~Leyva,
  arXiv:0905.4465 [hep-ph].
\cite{Peskin:1995ev}
\bibitem{Peskin:1995ev}
  M.~E.~Peskin and D.~V.~Schroeder,
{\it  Reading, USA: Addison-Wesley (1995) 842 p}
\bibitem{Blokland:2004nd}
  I.~R.~Blokland,
 ``Multiloop calculations in perturbative quantum field theory,''
 Alberta University thesis 2004, UMI-NQ-95909
\bibitem{Smirnov:1994tg}
  V.~A.~Smirnov,
  Mod.\ Phys.\ Lett.\  A {\bf 10} (1995) 1485
\bibitem{Smirnov:1996ng}
  V.~A.~Smirnov,
  Phys.\ Lett.\  B {\bf 394} (1997) 205
\bibitem{Beneke:1997zp}
  M.~Beneke and V.~A.~Smirnov,
  Nucl.\ Phys.\  B {\bf 522} (1998) 321

\bibitem{Tkachov:1981wb}
  F.~V.~Tkachov,
  Phys.\ Lett.\  B {\bf 100} (1981) 65.
\bibitem{Chetyrkin:1981qh}
  K.~G.~Chetyrkin and F.~V.~Tkachov,
  Nucl.\ Phys.\  B {\bf 192} (1981) 159.
\bibitem{Laporta:1996mq}
  S.~Laporta and E.~Remiddi,
  Phys.\ Lett.\  B {\bf 379} (1996) 283

\bibitem{Laporta:2001dd}
  S.~Laporta,
  Int.\ J.\ Mod.\ Phys.\  A {\bf 15} (2000) 5087

\bibitem{Fischer:2000kx}
  M.~Fischer, S.~Groote, J.~G.~K\"orner and M.~C.~Mauser,
  Phys.\ Rev.\  D {\bf 63} (2001) 031501
\bibitem{Czarnecki:1990kv}
  A.~Czarnecki,
  Phys.\ Lett.\  B {\bf 252} (1990) 467.
\bibitem{Piclum:2008zz}
  J.~H.~Piclum, A.~Czarnecki and J.~G.~K\"orner,
  Nucl.\ Phys.\ Proc.\ Suppl.\  {\bf 183} (2008) 48.
\bibitem{Frink:1997sg}
  A.~Frink, J.~G.~K\"orner and J.~B.~Tausk,
  ``Massive two-loop integrals and Higgs physics,''
  arXiv:hep-ph/9709490.
\bibitem{Brucher:1998ec}
  L.~Br\"ucher, J.~Franzkowski and D.~Kreimer,
  ``xloops: Automated Feynman diagram calculation,''
  Comput.\ Phys.\ Commun.\  {\bf 115} (1998) 140.
\bibitem{Bauer:2000cp}
  C.~W.~Bauer, A.~Frink and R.~Kreckel,
  ``Introduction to the GiNaC Framework for Symbolic Computation within 
  arXiv:cs/0004015.
\bibitem{Kadeer:2005aq}
  A.~Kadeer, J.~G.~K\"orner and U.~Moosbrugger,
  Eur.\ Phys.\ J.\  C {\bf 59} (2009) 27
\bibitem{Korner:1989ve}
  J.~G.~K\"orner and G.~A.~Schuler,
  Phys.\ Lett.\  B {\bf 231} (1989) 306.
\bibitem{Korner:1989qb}
  J.~G.~K\"orner and G.~A.~Schuler,
  Z.\ Phys.\  C {\bf 46} (1990) 93.
\bibitem{Faessler:2002ut}
  A.~Faessler, T.~Gutsche, M.~A.~Ivanov, J.~G.~K\"orner and V.~E.~Lyubovitskij,
  Eur.\ Phys.\ J.\ direct C {\bf 4} (2002) 18
\bibitem{Simmons:2000hr}
  E.~H.~Simmons,
  ``Top physics,''
  arXiv:hep-ph/0011244.
\bibitem{AguilarSaavedra:2007rs}
  J.~A.~Aguilar-Saavedra, J.~Carvalho, N.~F.~Castro, A.~Onofre and F.~Veloso,
  Eur.\ Phys.\ J.\  C {\bf 53} (2008) 689
\bibitem{Aaltonen:2008ei}
  T.~Aaltonen {\it et al.}  [CDF Collaboration],
  Phys.\ Lett.\  B {\bf 674} (2009) 160
\bibitem{d02008}
The D0 Collaboration, Model independent measurement of the W boson helicity in
top quark decays at D0, D0 note 5722-Conf, 20058 (2008).
\bibitem{Uhlemann:2008pm}
  C.~F.~Uhlemann and N.~Kauer,
  Nucl.\ Phys.\  B {\bf 814} (2009) 195
\bibitem{Han:2005mu}
  T.~Han,
  Lectures given at TASI 2004,
  ``Collider phenomenology: Basic knowledge and techniques,''
  arXiv:hep-ph/0508097.
\bibitem{Calderon:2008zz}
  G.~Calderon and G.~Lopez Castro,
  Int.\ J.\ Mod.\ Phys.\  A {\bf 23} (2008) 3525.
\bibitem{Larin:1993tq}
  S.~A.~Larin,
  Phys.\ Lett.\  B {\bf 303} (1993) 113

\bibitem{Larin:1991tj}
  S.~A.~Larin and J.~A.~M.~Vermaseren,
  Phys.\ Lett.\  B {\bf 259} (1991) 345.
\bibitem{pck09}
J.~H.~Piclum, A.~Czarnecki and J.~G.~K\"orner, to be published

\end{thebibliography}
%

\end{footnotesize}


\end{document}